\documentclass[preprintnumbers,nofootinbib,noshowpacs,eqsecnum,aps,superscriptaddress,letterpaper,twocolumn]{revtex4-2}

\usepackage{graphicx}
\usepackage{amsmath, amssymb}
\usepackage{xurl}
\usepackage{hyperref}
\usepackage{booktabs}
\usepackage{enumitem}
\hypersetup{
  colorlinks  = true,
  urlcolor    = blue,
  linkcolor   = blue,
  citecolor   = red
}
\usepackage[capitalize]{cleveref}

\newcommand{\rnu}{\langle r_\nu^2 \rangle}
\newcommand{\Eps}{\varepsilon}
\def\cevns{{CE$\nu$NS}}
\begin{document}

\title{Coherent Elastic Neutrino-Nucleus Scattering at the Japan Proton Accelerator Research Complex}

\author{J.I.\ Collar}
\email{collar@uchicago.edu}
\affiliation{Enrico Fermi Institute, Kavli Institute for Cosmological Physics, and Department of Physics\\
University of Chicago, Chicago, Illinois 60637, USA}
\affiliation{Donostia International Physics Center, \\ Paseo Manuel Lardizabal 4, 20018, Donostia-San Sebasti\'an, Spain}
\affiliation{Ikerbasque, Basque Foundation for Science, \\ Plaza Euskadi 5, 48013, Bilbao, Spain}

\author{Ivan Esteban}
\email{ivan.esteban@ehu.eus}
\affiliation{Department of Physics, University of the Basque Country UPV/EHU, PO Box 644, 48080
Bilbao, Spain}
\affiliation{EHU Quantum Center, University of the Basque Country UPV/EHU, PO Box 644, 48080
Bilbao, Spain}

\author{J.J. Gomez-Cadenas}
\affiliation{Donostia International Physics Center, \\ Paseo Manuel Lardizabal 4, 20018, Donostia-San Sebasti\'an, Spain}
\affiliation{Ikerbasque, Basque Foundation for Science, \\ Plaza Euskadi 5, 48013, Bilbao, Spain}

\author{M.~C.~Gonzalez--Garcia}
\email{concha.gonzalez-garcia@stonybrook.edu}
\affiliation{Departament  de  Fisica  Quantica  i  Astrofisica
 and  Institut  de  Ciencies  del  Cosmos,  Universitat
 de Barcelona, Diagonal 647, E-08028 Barcelona, Spain}
\affiliation{Instituci\'o Catalana de Recerca i Estudis Avancats (ICREA)
Pg. Lluis  Companys  23,  08010 Barcelona, Spain.}
\affiliation{C.N. Yang Institute for Theoretical Physics, Stony Brook University, Stony Brook NY11794-3849,  USA}

\author{L.-C. Ji}
\affiliation{Donostia International Physics Center, \\ Paseo Manuel Lardizabal 4, 20018, Donostia-San Sebasti\'an, Spain}

\author{L. Larizgoitia}
\affiliation{Donostia International Physics Center, \\ Paseo Manuel Lardizabal 4, 20018, Donostia-San Sebasti\'an, Spain}

\author{C.M.\ Lewis}
\affiliation{Enrico Fermi Institute, Kavli Institute for Cosmological Physics, and Department of Physics\\
University of Chicago, Chicago, Illinois 60637, USA}
\affiliation{Donostia International Physics Center, \\ Paseo Manuel Lardizabal 4, 20018, Donostia-San Sebasti\'an, Spain}

\author{F. Monrabal}
\affiliation{Donostia International Physics Center, \\ Paseo Manuel Lardizabal 4, 20018, Donostia-San Sebasti\'an, Spain}
\affiliation{Ikerbasque, Basque Foundation for Science, \\ Plaza Euskadi 5, 48013, Bilbao, Spain}

\author{Jo\~{a}o Paulo Pinheiro.}
\email{joaopaulo.pinheiro@fqa.ub.edu}
 \affiliation{State Key Laboratory of Dark Matter Physics, Tsung-Dao Lee Institute \& School of Physics and Astronomy, Shanghai Jiao Tong University, Shanghai 200240, China}
 \affiliation{Key Laboratory for Particle Astrophysics and Cosmology (MOE) \& Shanghai Key Laboratory for Particle Physics and Cosmology, Shanghai Jiao Tong University, Shanghai 200240, China}

\author{A. Sim\'on}
\affiliation{Donostia International Physics Center, \\ Paseo Manuel Lardizabal 4, 20018, Donostia-San Sebasti\'an, Spain}
\affiliation{Instituto de F\'isica Corpuscular, CSIC \& Universitat de Val\`encia, \\ Calle Catedr\'atico Jos\'e Beltr\'an 2, 46980, Paterna, Spain}

\author{S.G.\ Yoon}
\affiliation{Enrico Fermi Institute, Kavli Institute for Cosmological Physics, and Department of Physics\\
University of Chicago, Chicago, Illinois 60637, USA}

\date{\today}
\preprint{}

\begin{abstract}
The Japan Proton Accelerator Research Complex (J-PARC) currently
delivers a 1 MW, 3\,GeV proton beam to the Materials and Life Science
Experimental Facility (MLF). Power is expected to increase to 1.3 MW, driven by the needs
of Hyper-Kamiokande. As a result, the MLF presently provides the highest neutron
yield of any spallation source, while potentially holding the best current and foreseeable
conditions for Coherent Elastic Neutrino-Nucleus Scattering (CE$\nu$NS) experimentation.
We explore this potential, using as examples detector technologies presently funded for
construction and under development. We quantify their sensitivity to a rich variety of
particle physics scenarios, finding that very-high-statistics CE$\nu$NS measurements
with significant sensitivity to relevant scenarios are feasible at this facility within the next few years.
\end{abstract}

\maketitle


\section{Introduction}
\label{sec:intro}

The most probable interaction mechanism for low-energy  neutrinos is their elastic scattering off atomic nuclei, mediated by the neutral electroweak current. For neutrino energies below few tens of MeV the momentum exchange with
the recoiling nucleus is sufficiently low that the interaction must be
regarded as involving the nucleus as a whole. This coherent phenomenon
leads to a large enhancement to the elastic scattering cross section
of this process: it becomes proportional to the square of the number
of nucleons, although the numerical value of the weak mixing angle
makes the neutron contribution dominate~\cite{Freedman:1973yd}. This
process (Coherent Elastic Neutrino-Nucleus Scattering, CE$\nu$NS)
provides a new path to explore both neutrino properties and nuclear
structure. It also leads to a drastic reduction in neutrino detector
mass when compared to all other known means of neutrino interaction.

The experimental challenge resides in the fact that 
this process produces a single observable, a recoiling nucleus carrying a modest kinetic energy in the few-keV to sub-keV range, as is also the case in neutron elastic scattering and for the expected interactions from certain Dark Matter candidates.  For most targets only a small fraction of this energy is converted into a readily detectable form, e.g., ionization or scintillation: this energy-dependent conversion efficiency is commonly referred to as a “quenching factor” (QF). Its characterization is necessary for the interpretation of these faint neutrino signals. As a result, QF measurements presently constitute an active area of research involving numerous neutrino and Dark Matter detector materials.

Altogether this resulted in the four decades spanning between the theoretical description of CE$\nu$NS~\cite{Freedman:1973yd} and its experimental observation~\cite{COHERENT:2017ipa, Scholz:2017ldm} being consumed by a search for a viable combination of neutrino source and detector. The first was  provided by the Spallation Neutron Source (SNS) at Oak Ridge National Laboratory, the highest-power facility of this type at the time~\cite{Avignone:2003ep, Efremenko:2008an}. The second, by sodium-doped cesium iodide (CsI[Na]), an inorganic scintillator with ideal characteristics for this measurement~\cite{Collar:2014lya, nicole}. This long interval allowed for the development of a broad variety of CE$\nu$NS applications in nuclear and particle physics phenomenology:  searches for  electromagnetic properties of the neutrino, for non-standard neutrino-quark interactions facilitated by new mediators, studies of  weak nuclear charge and nuclear structure, evidence for sterile neutrinos and potential Dark Matter candidates, etc. An abridged list of the numerous publications in this area is provided in Sec.\ \ref{sec:pheno} of this work. The miniaturization that is possible with CE$\nu$NS detectors may also lead to eventual technological applications~\cite{Natti:2024esu}.

\begin{figure}[!htbp]
\includegraphics[width=.9 \linewidth]{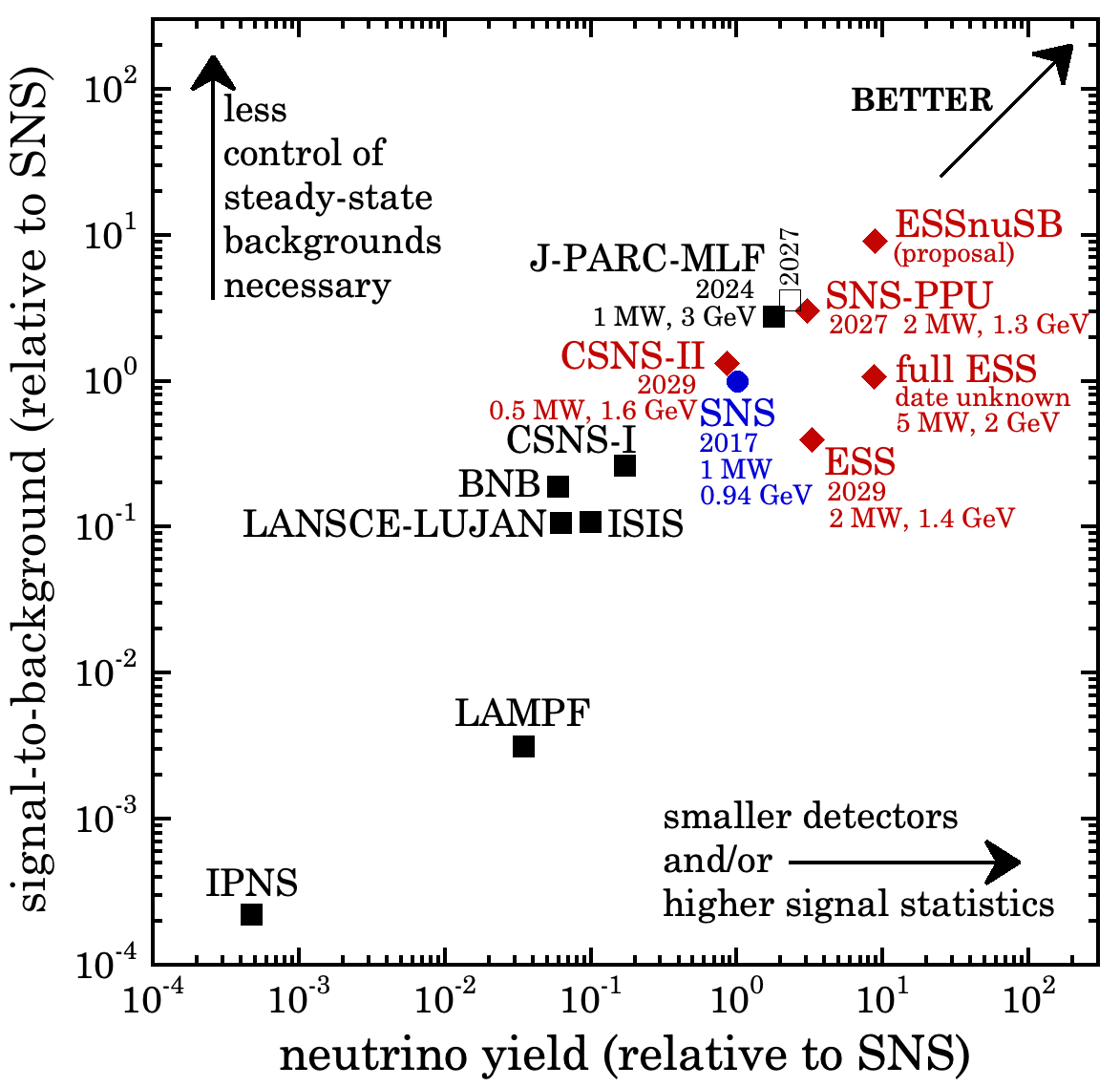}
\caption{\label{fig:facilities} {Comparison between past and present (black squares) and future (red diamonds) spallation sources from the perspective of CE$\nu$NS  detection, taking as  reference the SNS (blue circle) at the time of first observation of this process~\cite{COHERENT:2017ipa}. A hollow  square represents J-PARC MLF following the impending upgrade to 1.3 MW (see text).  }  }
\end{figure}

In a seminal paper, Drukier and Stodolsky~\cite{Drukier:1984vhf} examined the applicability of different neutrino sources to CE$\nu$NS  experimentation, concluding that spallation sources are the most immediately convenient. While those are mainly dedicated to neutron production, the decay-at-rest (DAR) of positive pions generated during the stopping of energetic protons in the spallation target also results in a high yield of three flavours of neutrinos ($\nu_{\mu}$, $\overline\nu_{\mu}$, $\nu_{e}$) having well-established spectral and timing characteristics as we discuss below~\cite{Avignone:2003ep, Efremenko:2008an}. All are able to participate in the coherent process described above. The pulsed nature of this flux helps reduce the impact of steady-state environmental backgrounds on CE$\nu$NS  detectors, and the relatively-high neutrino energy (up to the limit where coherence starts to  wane) results in easier-to-detect few-keV nuclear recoils. Lower-energy electron antineutrinos from nuclear power reactors provide a significantly higher flux in  proximity to their cores, yet without the benefit of a known time of arrival for CE$\nu$NS signals. Additionally, these reactor $\overline\nu_{e}$ generate harder-to-detect sub-keV recoils, for which the acquisition of a reliable knowledge of the QF is more of a challenge. To illustrate the difficulties involved, a first measurement of reactor CE$\nu$NS~\cite{Colaresi:2022obx} is now disputed in~\cite{Ackermann:2025obx}, with the adopted QF for sub-keV germanium recoils playing a critical role in resolving this tension~\cite{Li:2025pfw}. This situation will soon be clarified by upcoming results from the upgraded detector in \cite{Colaresi:2022obx}, now operating at the \mbox{Vandellòs-II} nuclear power plant, together with a novel precision measurement of the sub-keV QF in this medium. However, in view of it, it seems prudent to regard spallation facilities as the most reliable sources of CE$\nu$NS information, at least for the near future. The recent detection of CE$\nu$NS from comparatively low-flux solar neutrinos~\cite{XENON:2024ijk,PandaX:2024muv,LZ:2025igz} is to provide useful contributions~\cite{DeRomeri:2024iaw, AristizabalSierra:2024nwf} albeit limited by 
expected low statistics, 
even when  the multi-tonne  mass of the Dark Matter detectors involved is considered.  

Six years ago we described the prospects of the future European Spallation Source (ESS) for CE$\nu$NS work~\cite{Baxter:2019mcx}. In the interim, significant delays have affected the completion of the ESS. Whereas a design power of 5 MW and proton energy of 2 GeV was originally planned for 2027, the present goal has dwindled to 2 MW and 1.4 GeV in 2029, with additional delays possible given the novelty of its target design. This significantly impacts the immediate relevance of the ESS in this field, while the ESSnuSB expansion proposal~\cite{ESSnuSB:2023ogw} maintains great promise (Fig.~\ref{fig:facilities}). During this same period, the Japan Proton Accelerator Research Complex (J-PARC) has ramped up the power of its 3 GeV proton beam delivery to the Materials and Life Science Experimental Facility (MLF) to 1 MW, with an increase to 1.3 MW by 2027 being driven by the needs of the Hyper-Kamiokande detector. As a result, the MLF presently provides not only the highest neutron yield of any spallation source \cite{jparcn}, but also the best current and foreseeable conditions for CE$\nu$NS experimentation.

To illustrate this, Fig.~\ref{fig:facilities} (updated from~\cite{Abele:2022iml}) provides a balanced comparison between spallation sources in the context of CE$\nu$NS. The horizontal axis involves an explicit calculation of neutrino yield that includes the combined effect of proton energy, spallation target material, and beam current~\cite{Baxter:2019mcx}  instead of just beam power as in, e.g., Refs.~\cite{Kelly:2021jgj,scholberg}. The signal-to-background-ratio figure of merit along the vertical axis acknowledges the ability to assess ---and later subtract--- steady-state backgrounds during periods of anticoincidence with protons-on-target (POT), folding in the duty cycle of each source and assigning to all a common 10 $\mu$s post-POT window for the detection of  delayed  $\overline\nu_{\mu}$ and $\nu_{e}$ from $\mu^{+}$ decay following $\pi^{+}$ DAR~\cite{Abele:2022iml}. 

In this work, we analyze the potential of J-PARC MLF for CE$\nu$NS studies, using as examples detector technologies presently funded for construction and under development. We nevertheless consider an extrapolation to a future detector mass of order one tonne. We find that for  targets relatively straightforward to scale-up (e.g., cesium iodide) this represents a point of diminishing returns, one where the improvement in sensitivity to several parameters of phenomenological interest is no longer  limited by this factor (the exposure, as determined by detector mass) but instead is impacted by irreducible uncertainties associated to a realistically-achievable  knowledge of QF and neutrino flux.

The structure of this paper is as follows. Sec.\ \ref{sec:flux} describes the neutrino yield and signal structure expected at J-PARC MLF, listing additional advantages of this facility for CE$\nu$NS studies. Sec.\ \ref{sec:bckg_sources} describes the sources of background, both beam-related and environmental (i.e., steady-state) that would affect   detector technologies presented in Sec.\ \ref{sec:detectors}, during their deployment at the MLF. Background simulations derived from this information are used for the calculations of physics reach presented in Sec.\ \ref{sec:pheno}. Our conclusions are presented in Sec. \ref{sec:conclusions}.

\section{Neutrino yield at J-PARC MLF}
\label{sec:flux}

As described above, in a neutron spallation source neutrinos are mainly produced by the pion DAR $\pi^+ \rightarrow \mu^+ \, \nu_\mu$ followed by the muon DAR $\mu^+ \rightarrow e^+ \, \bar{\nu}_\mu\,\nu_e$~\cite{JSNS2:2017gzk}. The energy distribution is very-well understood: for  $\pi^+$ two-body decay neutrinos are monochromatic, for $\mu^+$ the SM provides a clean prediction~\cite{Avignone:2003ep, Efremenko:2008an, Breso-Pla:2025cul}. We provide analytic expressions in Sec.~\ref{sec:param}. The main uncertainty is the overall normalization, i.e., the number of pions produced per proton hitting the target. 

In this respect, an additional significant advantage of J-PARC MLF is the eventual ability of the  existing JSNS$^{2}$ detectors~\cite{JSNS2:2013jdh} to measure the local neutrino flux via the $^{12}C(\nu_{e},e^{-}) ^{12}N_{g.s.}$ reaction down to a 10\% uncertainty level~\cite{JSNS2:2013jdh}. A dedicated effort as has been proposed at other facilities~\cite{COHERENT:2021xhx} could reduce this further down to the 2\%--$5\%$ level.
Here, we 
assume a positive pion production per 3 GeV proton of 0.44 $\pi^{+}$/p at the MLF mercury target, derived from MCNPX  simulations~\cite{mcnpx} (ISABEL/Dresner intranuclear/evaporation models) like those described in Ref.~\cite{Baxter:2019mcx}. This is 
consistent with a recent JSNS$^{2}$ first {\it in situ} measurement of \mbox{0.48 $\pm$ 0.17}  $\pi^{+}$/p~\cite{JSNS2:2024uxo}. Our adopted pion production translates into a neutrino yield of $3.89 \times10^{22}~ \nu$ per flavour per year of continuous operation for 1.3 MW power delivery (the yearly operational schedule of MLF is included in Sec.\ \ref{sec:pheno}).  For perspective, this $\pi^{+}$/p production was $\sim$0.078 at the SNS at the time of the first CE$\nu$NS observation, a result of its marked dependence on proton energy~\cite{Baxter:2019mcx} (Fig.~\ref{fig:facilities}).

An additional advantage of J-PARC MLF, compared to the original prospects for the ESS~\cite{Baxter:2019mcx}, is the pulsed nature of the proton beam: 25 Hz of 0.1\,$\mu$s-wide double pulses separated by 0.54\,$\mu$s~\cite{JSNS2:2017gzk}. As a consequence, $\nu_\mu$ from $\pi^+$ decay, produced simultaneously with protons hitting the target due to the short pion lifetime $\tau_\pi \simeq 0.03\,\mu\mathrm{s}$ can be well-separated from $\bar{\nu}_e$ and $\bar{\nu}_\mu$ from $\mu^+$, produced after timescales of order the muon lifetime $\tau_\mu \simeq 2\, \mu\mathrm{s}$. This significantly boosts physics searches for flavour-dependent effects, that can exploit the time structure of the neutrino signal irrespective of flux normalization uncertainties. We provide explicit expressions and illustrate this further in Sec.~\ref{sec:param}. 

\section{Background sources affecting a CE$\nu$NS search}
\label{sec:bckg_sources}

Neutrino signals from CE$\nu$NS at an spallation source present a characteristic structure in both energy and time~\cite{Avignone:2003ep, Efremenko:2008an, Baxter:2019mcx}.  This facilitated the first experimental observation of this process~\cite{COHERENT:2017ipa, Scholz:2017ldm}. The time dependence, which closely traces the periodic injection of short proton spills into the spallation target, allows to reduce the impact of the environmental (i.e., steady-state) backgrounds that continuously affect  detectors operated without the benefit of a significant overburden. This reduction is achieved by selecting narrow coincidence windows following POT. Including a 10 $\mu$s allowance for the arrival of delayed  $\overline\nu_{\mu}$ and $\nu_{e}$  (Sec.\ \ref{sec:intro}) this background reduction is by a factor of 2.5$\times 10^{-4}$ at J-PARC MLF. This derives from the beam duty cycle described in the previous section. Additionally, steady-state backgrounds can be characterized at times preceding POT. This allows their subtraction from the energy spectrum of POT-coincident signals, leaving only the contributions from CE$\nu$NS and any beam-related backgrounds. Following their dedicated {\it in situ} characterization previous to CE$\nu$NS experimentation \cite{COHERENT:2017ipa, Scholz:2017ldm}, beam-related backgrounds can be diminished to a negligible level  by careful shielding of the detectors. For instance, a fraction of highly-energetic (hundreds of MeV) prompt  neutrons can escape the  shielding monolith around the spallation target to reach the detector within POT-coincident windows. Even if the time structure of their signals is not identical to that expected from CE$\nu$NS, proper shielding and/or vetoing of the background they induce is of the essence. In order to evaluate the potential at J-PARC of the detector technologies described in the next section, Geant4~\cite{GEANT4:2002zbu} and MCNPX~\cite{mcnpx} simulations were used to compare various expected background sources with the calculated rate of \cevns\ events. The backgrounds considered and the specific input to the simulations are as follows:

\begin{enumerate}[wide, labelwidth=!, labelindent=0pt]

    \item \textbf{environmental neutrons}

    Environmental neutrons (cosmic-ray tertiaries and those from fission decays and ($\alpha,n$) reactions) continuously affect detectors not benefiting from a significant overburden as is the case for experimental sites in close vicinity to spallation sources. We adopt their spectral hardness as described in~\cite{Gordon:2004non, env_neutrons_2}. Their thermal and epithermal fluxes do not appreciably contribute to the backgrounds in the devices considered here, for the shielding assumed. Fast neutrons (0.1 - 20 MeV) are simulated assuming an isotropic origin. An ultra fast component (20 MeV - 1 GeV) is simulated with an origin biased skyward~\cite{Sato:2016eyu}. The supplementary  flux via ($n$,X$n$) reactions that energetic neutrons can produce in shielding (primarily in lead) is included in our simulations.
    
    \item \textbf{internal radiopurity}

    The intrinsic radioactive backgrounds affecting each detector technology are simulated using Geant4~\cite{GEANT4:2002zbu}. PPC germanium detectors are the exception, as this steady-state background is available to us as a measurement, in conditions of negligible overburden as expected at J-PARC MLF. Contributions from the full decay chains of the various radioactive impurities are scaled by activities derived from the screening of detector materials. Those contributions are reduced by suitable cuts (e.g., coincidences between multiple detector elements or with active vetoes, as in the case for CsI).

    \item \textbf{muon-induced neutrons}

    Muons at Earth (cosmic-ray secondaries) traversing the detector and shielding geometries engender tertiary neutrons that can in turn produce nuclear recoil events in the same energy region as CE$\nu$NS signals. These neutrons are simulated as homogeneously produced in the lead shielding of each detector, with a hardness following \cite{Malgin:2015lna, Kluck:2013xga}. Their contribution is diminished by the measured tagging efficiency (typically $>99.9\%$) of an external muon veto surrounding the geometry.
    
    \item \textbf{beam-related neutrons}

    As mentioned, directly competing with the \cevns\ signal are POT-coincident neutron-induced signals from energetic neutrons escaping the shielding monolith around the spallation target. Taking the overall neutron flux measured at the SNS as a reference~\cite{COHERENT:2017ipa,Scholz:2017ldm}, scaled up for J-PARC's expected power delivery of 1.3 MW and the higher yield of neutrons from 3 GeV protons, simulations similar  to those described in~\cite{Collar:2014lya, Lewis:2023sbl} were performed to find the  background those induce in the detectors considered. A power law approximation as  in~\cite{jsns2_2,Ajimura_2015} describes the hardness of their spectrum in the third floor terrace of the MLF.

    \item \textbf{neutrino-induced neutrons (NINs)}

    The second directly competing signal to \cevns\ arises from the charged-current interaction $^{208}$Pb$(\nu_{e},e^{-}Xn)$ in the lead surrounding each detector. Homogeneously distributed origins and an isotropic emission are adopted, with an spectral hardness derived from \cite{Gardiner:2021qfr}, which is nearly identical to that used in \cite{COHERENT:2017ipa,Scholz:2017ldm}. The neutron production rate from \cite{Lazauskas:2010rh}  is scaled to the $\sim2.58\times10^{7}$ $\nu_{e}$/cm$^{2}$/s  expected 20 m away from the J-PARC MLF mercury target, with a  multiplicative correction by a factor of 0.29  applied to account for the upper limit to this reaction imposed in \cite{COHERENT:2022fic}.

\end{enumerate}

\section{Detector technologies}
\label{sec:detectors}

\subsection{Cryogenic CsI}

\begin{figure}[!htbp]
\centering
\includegraphics[width=1. \linewidth]{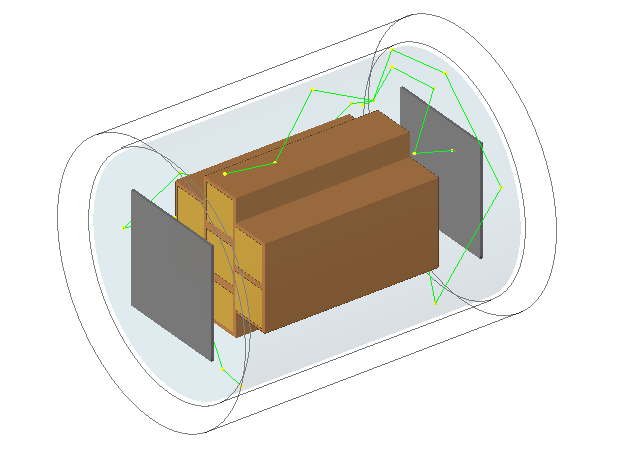}
\caption{\label{fig:COSI} { Rendition of the Geant4 geometry for the cryogenic CsI detector. Seven CsI crystals within OFHC copper holders, the LAr volume around them (gray), and two (dark gray) 20 $\times$ 20 cm SiPM tiles \cite{sipm} are visible. Also shown is an example track by a LAr scintillation photon originating from a  background event,  wavelength-shifted and reflected towards the  SiPM panels (see text). }  }
\end{figure}

This detector, conceptually described in \cite{Baxter:2019mcx} and depicted in Fig.\ \ref{fig:COSI}, is a compact 44 kg array of seven cryogenically-operated pure (i.e., undoped)  CsI crystals submerged in a liquid argon (LAr) bath. Similar designs using this same scintillating material have been proposed by others \cite{cohcsi,clovers}. The QF for this material in the energy region relevant to CE$\nu$NS has been recently measured \cite{Lewis:2021cjv, csinew}. Each crystal has dimensions of $6.6\times6.6\times32$ cm. A low-noise, high-efficiency large area avalanche photodiode (LAAPD, \cite{laapd1}) on each end reads out the scintillation light produced by radiation interactions in the CsI scintillator. The LAr bath doubles as thermal control ---maintaining temperature in the regime that provides a maximum light yield  from this material \cite{Lewis:2021cjv} and best LAAPD performance \cite{csinew}--- and as an active internal veto against neutrons able to penetrate the passive shielding surrounding the detector. This consists  of 15 cm of Pb, 30 cm of polyethylene, and 5 cm of plastic scintillator (muon veto). 

The modular segmentation of the CsI crystal array  assists in background event rejection, whether beam-related or steady-state, via coincidence cuts. For purposes of this simulation, a layer of tetraphenyl butadiene (TPB) waveshifter coating a Tetratex reflector \cite{tetra} is assumed to cover all surfaces inner to the LAr volume.  Its conversion efficiency and emission properties~\cite{Araujo:2019nag} were used for optical simulations mapping the  position-dependent efficiency of this inner veto to wave-shift and transport LAr scintillation light into two submerged large-area silicon photomultiplier (SiPM) arrays. SiPMs like those described at \cite{Rogers:2024cvk,sipm} have a sufficiently-low dark count rate in cryogenic conditions to allow operation of two 20 $\times$ 20 cm tiles like those in Fig.\ \ref{fig:COSI} at single photoelectron sensitivity, while causing only a modest (few percent) dead time.

\begin{table}[!htbp]
\centering
\begin{tabular}{ c|c|c| }
 \cline{2-3}
  & Amcrys CsI[Na] & SICCAS CsI \\
 \cline{2-3}
 Th-232 & $<0.5$ & 0.03 \\
 U-238 & 2.4 & 0.09 \\
 K-40 & 16.7 & $<4.1$ \\ 
 Cs-137 & 27.9 & 1.3 \\ 
 Cs-134 & 25.9 & 33 \\
 Rb-87 & 38 ppb & 1.8 ppb \\
 \cline{2-3}
\end{tabular}
\caption{ Radiopurity of the CsI[Na] scintillator used for \cite{COHERENT:2017ipa, Scholz:2017ldm} and the SICCAS stock being considered for the cryogenic CsI array (units of mBq/kg except for Rb-87). Cs-134 is a result of thermal neutron activation, i.e., expected to be invariant at equilibrium. } \label{tbl:radio_csi}
\end{table}

Order-of-magnitude improvements in  the radiopurity of commercially-available CsI stock from SICCAS~\cite{siccas} (Table I) results in a background contribution of just $\sim30$ counts/keV-kg-day from this internal source before the reduction derived from beam duty-time. Measurements made by the manufacturer and at SNOLAB's Low Background Counting Facility, shown in Table \ref{tbl:radio_csi}, illustrate the improvements fed into present simulations.

\begin{figure*}[!hbtp]
\centering
\includegraphics[width=0.9\textwidth]{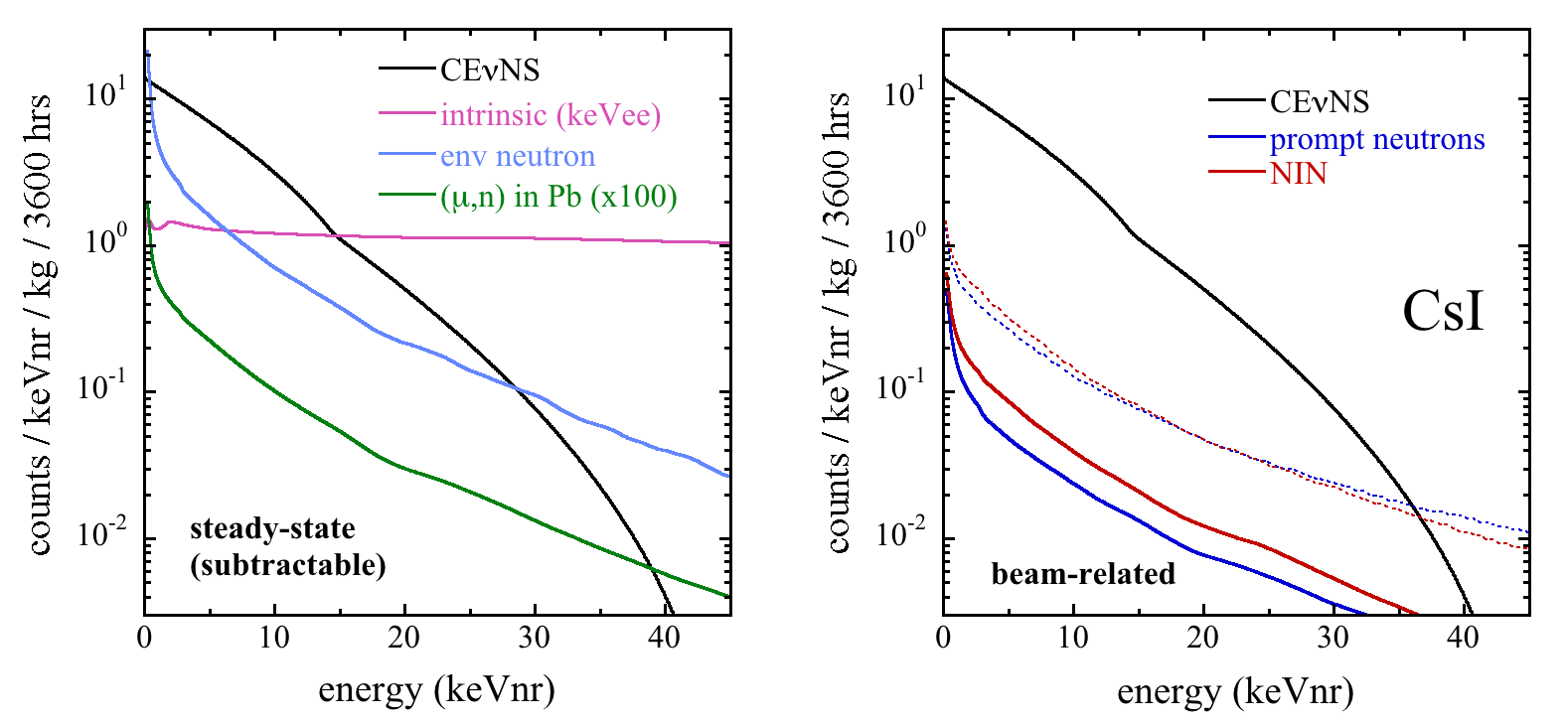}
    \caption{\label{fig:induced_spectra_csi} { {\it{Left:}} Subtractable steady-state backgrounds induced in the CsI detector array alongside the expected \cevns\ rate. The reduction in magnitude for these backgrounds due to beam duty time at J-PARC MLF is  applied here. To convert the intrinsic background to keVnr, we assume an 8\% QF~\cite{Lewis:2021cjv}.  {\it{Right:}} Beam-associated backgrounds (non-subtractable) in comparison to the expected \cevns\ signal. Dotted lines indicate their magnitude prior to applying  LAr inner veto cuts.}  }
\end{figure*}

Fig.~\ref{fig:induced_spectra_csi} aggregates the contributions of the various background sources simulated, comparing them to the expected \cevns\ rate. The steady-state backgrounds shown in the left panel of Fig. \ref{fig:induced_spectra_csi} can be subtracted via their characterization during the 10 $\mu$s prior to the POT trigger \cite{COHERENT:2017ipa, Scholz:2017ldm}.  Beam-coincident backgrounds are rendered  subdominant to the CE$\nu$NS signal with the active and passive shielding layers assumed here. 

We adopt a conservative energy threshold of 4.5 keVnr for this detector, which derives from the rapid decrease in the low-energy QF observed in Ref.~\cite{Lewis:2021cjv}.

\subsection{Ge PPC}

\begin{figure}[!htbp]
\centering
\includegraphics[width=.9 \linewidth]{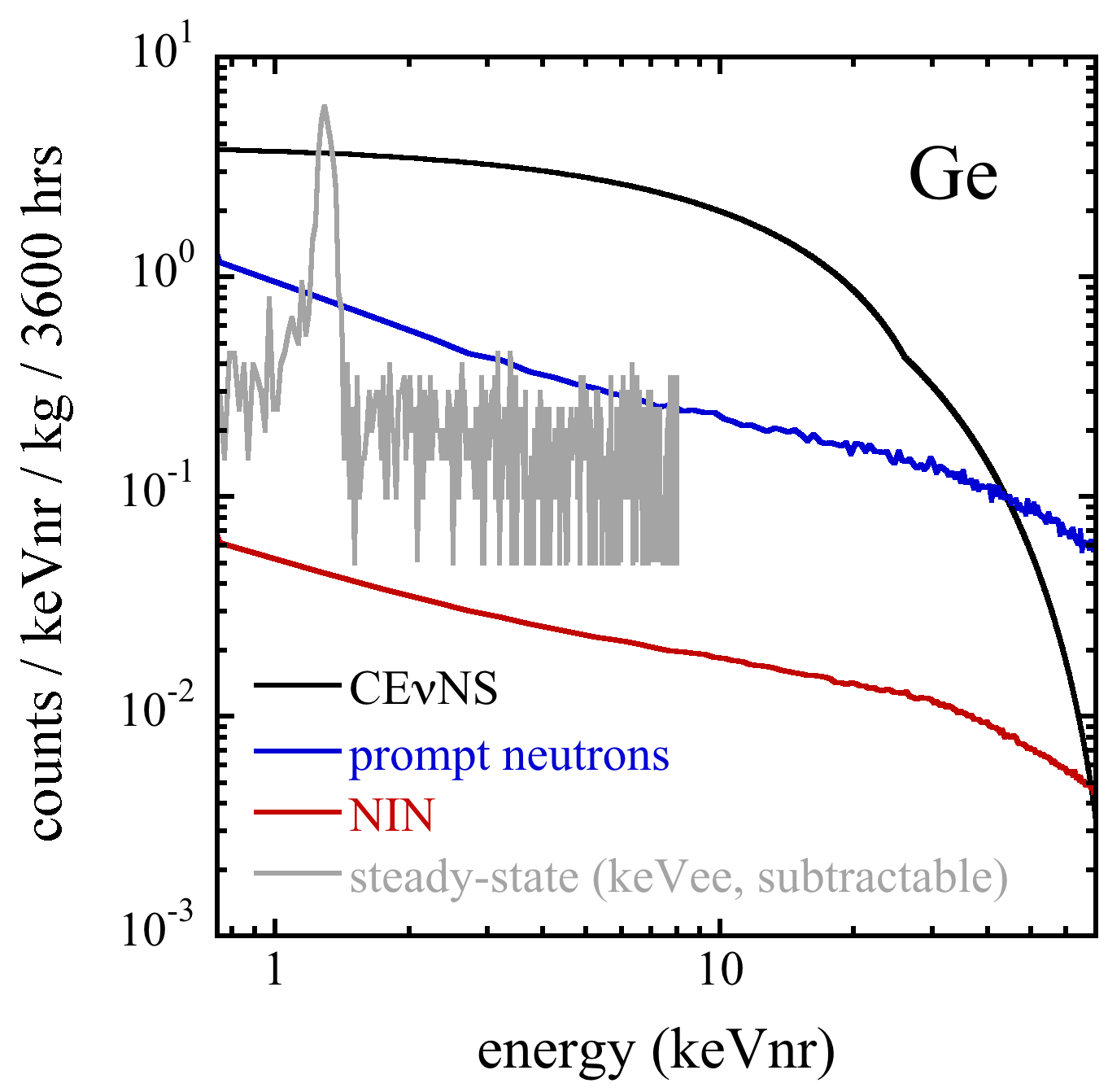}
\caption{\label{fig:GeBckgr} { Beam-associated (non-subtractable) and subtractable steady-state backgrounds in comparison to the expected \cevns\ signal in the Ge detector (see text). The steady-state component is reduced by the beam duty time and expressed in per keV$_{ee}$ units. For a more direct comparison, the rest of spectra can be converted to the same energy scale using the QF for germanium, that we set to 20\% in our simulations~\cite{geqf2}. } }
\end{figure}

A large-mass (3 kg) p-type point-contact (PPC, \cite{ppc})  germanium detector considered here is an upgraded design currently in operation at the \mbox{Vandellòs-II} nuclear power plant. Its shielding structure is identical to that described in~\cite{Colaresi:2021kus}, consisting of a 5 cm inner  active veto (plastic scintillator), 15 cm of Pb (innermost 2.5 cm low in Pb-210), and a 5 cm outer muon veto. Attention has been paid to the radiopurity and cleaning of internal PPC components and those in the inner veto, achieving a background level of 25 counts/keV-kg-day at 0.2 keV$_{ee}$  under a shallow 6 m.w.e.\ overburden \cite{Baxter:2019mcx} (``ee" refers to ``electron equivalent", i.e., ionization energy). This available measurement of the steady-state background takes the place of a dedicated  simulation for purposes of present estimates.

Fig.~\ref{fig:GeBckgr} shows the background in the Ge detector, simulated and measured, in comparison to the expected \cevns\ rate. Roughly an order of magnitude higher signal than background rate is expected at threshold. The ease of the addition of an external neutron moderator such as polyethylene, not included in the present simulation, makes this a conservative estimate of the prompt neutron-induced background. The steady-state backgrounds are subtractable by comparing data coincident with beam spills with data in the periods prior, as for CsI.

We adopt an energy threshold of 0.3 keVnr. This derives from the lowest-detectable ionization signal in a modern PPC benefiting from an external trigger, when a 20\% QF is adopted as a compromise between the values postulated in Refs.~\cite{Colaresi:2022obx, Ackermann:2025obx}.

\subsection{Gaseous TPC}
\label{subsec:gas_tpc}

\begin{figure*}[!hbtp]
\centering
\includegraphics[width=0.65\textwidth]{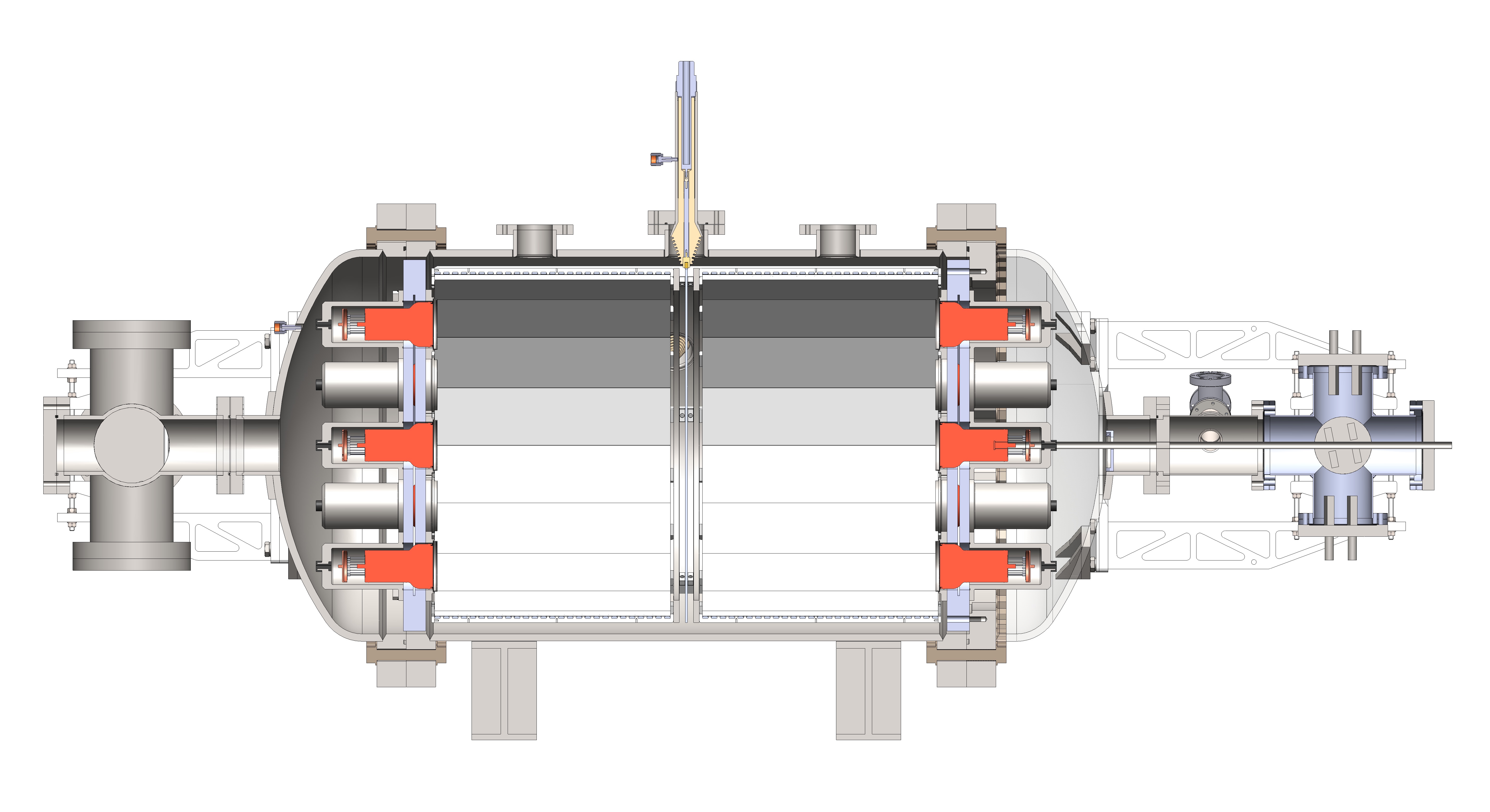}
\caption{\label{fig:ganess_detector} Lateral cut of the GanESS detector design, consisting of two symmetrical TPCs with the cathode in the center of a stainless steel 316Ti alloy vessel, allowing for operation above 35 bar. Two planes of 19 PMTs each read both the primary scintillation light produced by the events and the amplified signal of the ionised electrons when crossing the electroluminescent regions in the laterals of the detector. This detector  allows for operation with any noble gas, in particular argon and xenon which are the gases assumed in this work.}
\end{figure*}

Noble gases Time Projection Chambers (TPCs) with electroluminescent amplification is a technology that has revolutionized Dark Matter searches in the last 20 years by allowing for a large exposure with high sensitivity to low energy depositions in the active volume thanks to its intrinsic signal amplification. In this case the GanESS collaboration proposes to use a single-phase high-pressure gas electroluminescent TPC providing the capability to change between different noble gases operation with no modification of the detector. The high pressure gas TPC is a technology that has been largely developed for neutrinoless double beta decay searches within the NEXT Collaboration~\cite{NEXT:2025yqw} and can be extended to low energy searches. In addition, the use of gas TPC has the advantage of allowing for signal amplification without the issues of ``spontaneous'' single electron signals that often follow a large event that limits the surface operation of dual-phase detectors. Also, in comparison with single phase detectors, the amplification capabilities will allow for an improved energy threshold. For concreteness, we assume the same energy threshold and QF as in Ref.~\cite{Baxter:2019mcx}.

Fig.~\ref{fig:ganess_detector} shows a transverse cut of the design of the GanESS detector, with two symmetric TPCs of 30 cm drift length and 60 cm diameter with a common cathode. The two TPCs are placed inside a stainless steel 316Ti radio-pure alloy that allows the detector to operate above 35 bar. 
Two electroluminescent regions are responsible for the signal amplification in each side of the detector. That signal is read by a plane of 19 Hamamatsu R11410-20 3 inch PMTs just behind the amplification region.
These PMTs are protected from the pressure volume by an 8 mm thick sapphire window sealed with a steel frame. The windows are coated with a resistive polymer (PEDOT) to guarantee proper grounding on the surface and prevent any field leackage into the PMT. On top of the polymer, we added a coating of tetraphenyl butadiene (TPB) to transform the VUV light emitted by noble gases into blue light, for which sapphire is not opaque. In addition, the inner walls of the TPCs are covered by teflon panels also coated with TPB to maximize light collection.
The GanESS detector will use the vessel and PMTs employed for the NEXT-White detector~\cite{NEXT:2018rgj} where all materials were screened and characterized for neutrinoless double-beta decay searches~\cite{NEXT:2019rum}.

\begin{figure}[!htbp]
\centering
\includegraphics[width=.9 \linewidth]{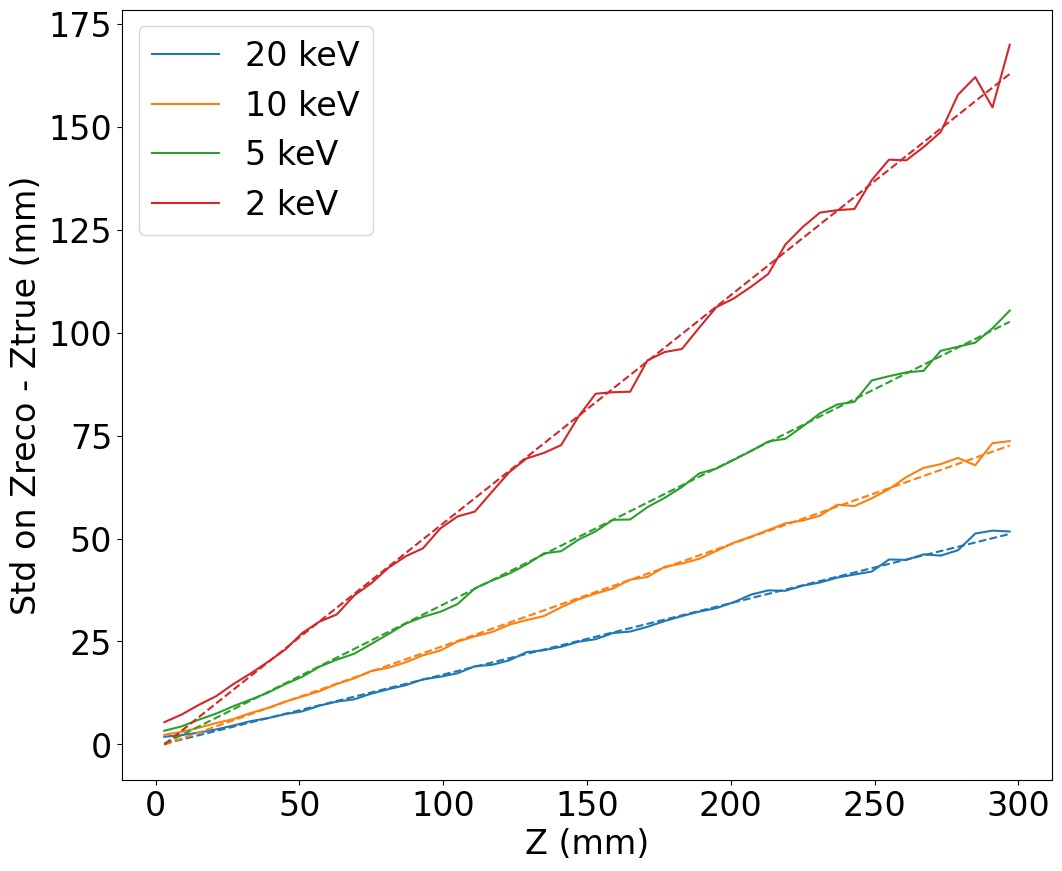}
\includegraphics[width=.9 \linewidth]{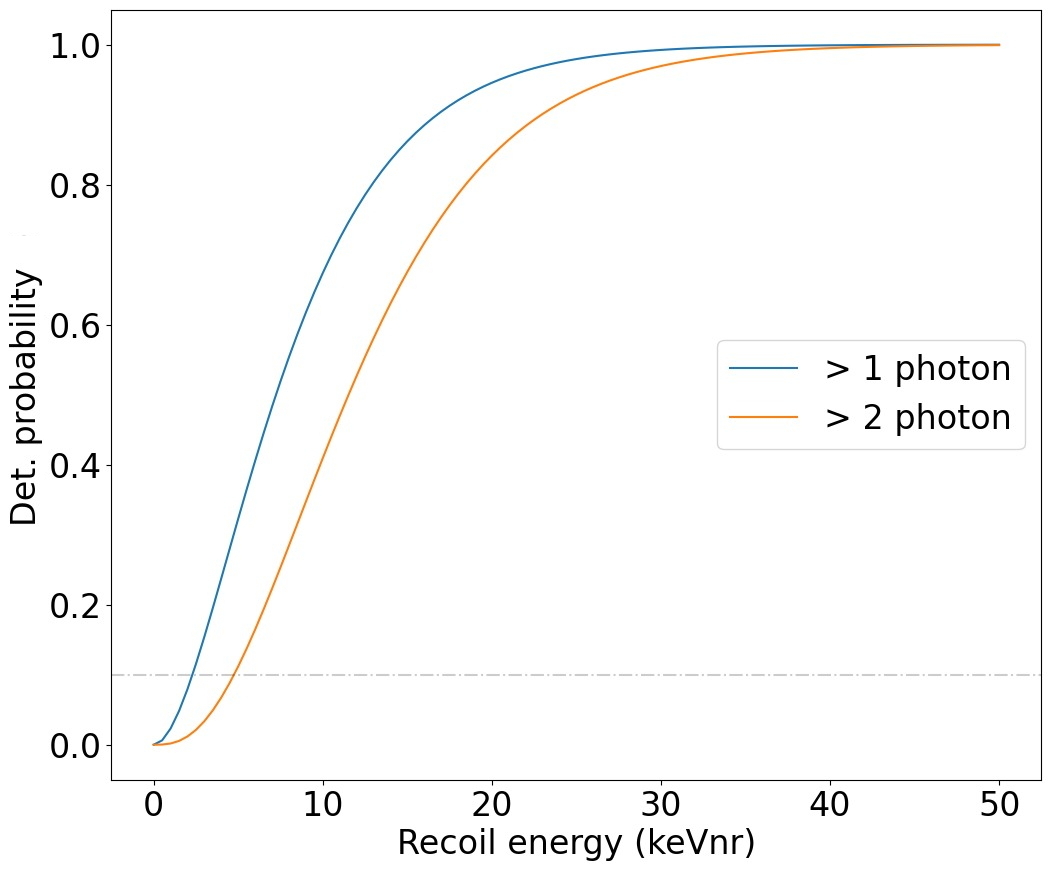}
\caption{\label{fig:ganees_timeres} {Top: Error on the reconstruction of the z-position of events at different energies as a function of their distance to the amplification region for the case of a high-pressure gaseous xenon detector (see text). Bottom: Detection probability of the primary scintillation light of more than 1 or 2 photons as a function of the nuclear recoil energy also calculated for xenon operation. A horizontal line represents 10\% detection probability.}}
\end{figure}

In this technology, the time resolution for the amplified signal is limited to the total time drift of the whole active volume ($\sim$300 $\mu$s in this case). If the drift time can be determined, the time resolution can then be of only a few nanoseconds. There are two different possibilities for determining the total drift distance: detection of the primary scintillation signal and measuring the electron diffusion, which is proportional to the square root of the drift time. In the case of very low energy signals, measuring the diffusion is complicated due to the limited number of ionised electrons. Fig.\ \ref{fig:ganees_timeres} shows the error on the estimation of the z-position as a function of the interaction distance to the amplification region for different recoil energies. It can be observed that this method is not particularly good as the error becomes larger than tens of millimetres in almost any situation, which will translate into tens of microseconds. On the other hand, for a system operated with PMTs with negligible dark count, identification of single photons is possible. In this case, the observation of a signal of 1 or 2 photons in the PMT plane in the time window between the arrival of the pulse and the amplified signal will be associated to the detection of the primary scintillation signal. Fig.\ \ref{fig:ganees_timeres} bottom shows the detection probability of more than one and more than 2 photons in a time window of 100 ns as a function of the nuclear recoil energy. As can be seen in this plot, we obtain moderate efficiencies even at low energies. In addition, in light of the relevance of the time information for the physics case of this process (see Sec.~\ref{sec:pheno}), the detector could be designed for an improved collection of the primary signal.

The main backgrounds affecting this technology are the ones described in Sec.~\ref{sec:bckg_sources} with an important difference between the operation with xenon and argon, as natural argon includes the $^{39}$Ar isotope that introduces a constant and flat background in the region of interest at the level of 14.6 evt/keV/kg/year, which is the dominant contribution coming from inner components. 
Initial simulations of the background events produced by  beam-related neutrons include two layers of passive shielding, 30 cm of polyethylene with an inner layer of 15 cm of lead. These simulations produce an almost flat spectrum at the level of 2 evt/keV/kg/year. The time dependence of this background has also been implemented taking into account the time profile of the J-PARC beam with two pulses separated by 540 ns.


\section{Physics Reach}
\label{sec:pheno}
A precision measurement of CE$\nu$NS provides a direct probe of both
SM and beyond the standard model (BSM) physics. Paradigmatic examples
of the former are the determination of the weak mixing angle at very
low momentum transfer~\cite{Canas:2018rng, Cadeddu:2018izq,Huang:2019ene,Cadeddu:2019eta, Cadeddu:2020lky, Cadeddu:2021ijh,AtzoriCorona:2023ktl,AtzoriCorona:2025xgj} and the study of nuclear structure~\cite{Cadeddu:2017etk,Ciuffoli:2018qem, Papoulias:2019lfi, Cadeddu:2021ijh,Coloma:2020nhf,AtzoriCorona:2023ktl,AtzoriCorona:2025xgj}.
The program of BSM exploration with CE$\nu$NS is
broad (see, e.g., Refs.~\cite{Barranco:2005yy, Formaggio:2011jt,Anderson:2012pn, Dutta:2015nlo, Cerdeno:2016sfi, Dent:2016wcr,Coloma:2017egw, Kosmas:2017zbh, Ge:2017mcq, Shoemaker:2017lzs,Coloma:2017ncl, Liao:2017uzy, Canas:2017umu, Dent:2017mpr,Papoulias:2017qdn, Farzan:2018gtr, Billard:2018jnl, Coloma:2019mbs,Chaves:2021pey, AristizabalSierra:2018eqm, Brdar:2018qqj,Cadeddu:2018dux, Blanco:2019vyp, Dutta:2019eml, Miranda:2019wdy,CONNIE:2019swq, Dutta:2019nbn, Papoulias:2019txv, Khan:2019cvi,Cadeddu:2019eta, Giunti:2019xpr, Baxter:2019mcx, Canas:2019fjw,Miranda:2020zji, Flores:2020lji, Miranda:2020tif, Hurtado:2020vlj,Miranda:2020syh, Cadeddu:2020nbr, Shoemaker:2021hvm,delaVega:2021wpx, Liao:2021yog, CONUS:2021dwh, Flores:2021kzl,Li:2022jfl, AristizabalSierra:2019zmy, Abdullah:2020iiv,Fernandez-Moroni:2021nap, Bertuzzo:2021opb, Bonet:2022imz,DeRomeri:2022twg,Coloma:2022avw} for an
incomplete list), being most sensitive to a variety of scenarios
leading to modified neutrino interactions with nuclei --- in particular
at low momentum transfer --- but extending also to the production of
new light neutral states and sterile neutrino searches, among others.

In what follows we illustrate, for a variety of target
nuclei and detection technologies,
the potential sensitivity of a CE$\nu$NS experiment at J-PARC
to SM and BSM physics: the weak mixing angle, neutrino charge radius, neutron radius, 
non-standard neutrino interactions (NSI), new neutral interactions with light-mediators, anomalous neutrino magnetic moment,
and eV-scale sterile neutrinos.

\subsection{Parameters and assumptions used in the calculations}
\label{sec:param}
\begin{table*}[!htbp]
  \renewcommand{\arraystretch}{1.4} \centering
  \begin{tabular}{|@{\hspace*{2pt}}c@{\hspace*{2pt}}|@{\hspace*{2pt}}c@{\hspace*{2pt}}|@{\hspace*{2pt}}c@{\hspace*{2pt}}|@{\hspace*{2pt}}c@{\hspace*{2pt}}|@{\hspace*{2pt}}c@{\hspace*{0pt}}|@{\hspace*{2pt}}c@{\hspace*{2pt}}|@{\hspace*{2pt}}c@{\hspace*{2pt}}|@{\hspace*{2pt}}c@{\hspace*{2pt}}|@{\hspace*{2pt}}c@{\hspace*{2pt}}|@{\hspace*{2pt}}c@{\hspace*{2pt}}|
    }
    \hline Detector Technology & Target & Mass & Energy Thres.  & Energy Resol.
    & Time Resol. & CE$\nu$NS  & BR bck 
    & NiN bck  & SS bck 
    \\[-0.1cm]
                              & nucleus & (kg) & ($\equiv T_{\rm th}$, keV$_{\rm nr}$) 
    & at $T_{\rm th}$ ($\equiv \sigma_{T,0}$, \%)  &  ($\equiv \sigma_t$) & $\frac{\rm NR}{\rm yr}$
    &$\frac{\rm NR}{\rm yr}$  & $\frac{\rm NR}{\rm yr}$ & $\frac{\rm NR}{\rm yr}$ \\ \hline
Cryogenic scintillator   & CsI & 44 & 4.5 & 30 & 0.7234 $\ln$(ns) & 1296 & 12 & 25 & 961\\
p-type point contact HPGe & Ge & 7 & 0.3 & 15 & 85 ns & 225 & 64 & 3.8 & 33 \\
High-pressure gaseous TPC & Xe & 20 & 0.9 & 40  &
\begin{tabular}{r|c}S1 events & 100 ns\\[-0.1cm]
  NoS1 events & --\end{tabular}
  &
  \begin{tabular}{c} 253\\[-0.1cm]  823 \end{tabular}
  &
  \begin{tabular}{c} 25\\[-0.1cm]  39 \end{tabular}
  &
  negl. 
  &
  negl. 
    \\
    High-pressure gaseous TPC & Ar & 6 & 0.9 & 40 &
\begin{tabular}{r|c}S1 events & 100 ns\\[-0.1cm]
  NoS1 events & --\end{tabular}
  &
  \begin{tabular}{c} 80\\[-0.1cm]  40 \end{tabular}
  &
  \begin{tabular}{c} 69\\[-0.1cm]  99 \end{tabular}
  &
  negl.
  &
  \begin{tabular}{c} 623\\[-0.1cm]  894 \end{tabular}
    \\ \hline
  \end{tabular}
\caption{\label{tab:detectors} Summary of detector properties
          used in our sensitivity calculations.}
\end{table*}

All sensitivity calculations in this work are obtained by binning the observed
events in reconstructed nuclear recoil energy, $T^\mathrm{r}$
(after accounting for the effect of the QF on the detector threshold and 
backgrounds, and taking the QF uncertainty into account) and
reconstructed time, $t^\mathrm{r}$.
Table~\ref{tab:detectors} summarizes the detector properties assumed in the simulation.
For the HPGe (high-purity germanium) PPC and Cryogenic-CsI detectors we assume
that time can be reconstructed for all
events. For TPCs, we distinguish two event samples. On the one hand, {\sl S1 events},
where the prompt scintillation light is detected, allowing to reconstruct the interaction
time with good accuracy as described in Sec.~\ref{subsec:gas_tpc}. These occur with probability $\epsilon_t(T)$, obtained from detector
simulation (see Sec.~\ref{subsec:gas_tpc} and Fig.~\ref{fig:ganees_timeres}). We conservatively include only events 
with more than 2 scintillation photons. This suppresses the number of events with time discrimination at low recoil energies. On the other hand, {\sl NoS1 events}, where the 
prompt scintillation light is 
not detected and for which the interaction time cannot be constructed. 
These occur with probability $1 - \epsilon_t(T)$.

The double differential predicted number of signal events in the reconstructed variables reads
\begin{equation}
 \!\! \frac{\mathrm{d}^2N^\mathrm{S}}{\mathrm{d}T^\mathrm{r}\, \mathrm{d} t^\mathrm{r}}=\!
 \int_{0}^{t_{\rm max}}
 \!\!\!\!\!\! \!\!\!\mathrm{d} t \int_{T_\mathrm{th}}^\infty \!\!\!\!\!\mathrm{d}T \,
  \frac{\mathrm{d}^2 N^\mathrm{S}}{\mathrm{d}t \,\mathrm{d}T} R_T(T^\mathrm{r}, T)
  R_t (t^\mathrm{r}, t) \epsilon_t(T)\:,
  \label{eq:doublerecdist}
\end{equation}
where  $\frac{\mathrm{d}^2 N}{\mathrm{d}t\,\mathrm{d}T}$ is the distribution of events
in true time, $t$ (we set $t=0$ when protons hit the target); and true nuclear recoil 
energy, $T$. $R_T(T^\mathrm{r}, T)$ and
$R_t (t^\mathrm{r}, t)$ are the corresponding energy and time resolution functions. 
$\epsilon_t(T)=1$ for HPGe PPC and Cryogenic-CsI. For TPCs the above expression is valid
for S1 events, while for NoS1 events we only bin in reconstructed energy 
\begin{equation}
\!\!  \frac{dN^\mathrm{S,NoS1}}{dT^\mathrm{r}}\!=\!
\int_{0}^{t_{\rm max}} \!\!\!\!\!\!\!\!\!\! d t \int_{T_\mathrm{th}}^\infty \!\!\!\!\!\!dT \,
  \frac{d^2 N^\mathrm{S}}{dt dT} R_T(T^\mathrm{r}, T) [1-\epsilon_t(T)]\;.
  \label{eq:noS1dist}
\end{equation}

For all detectors we assume Gaussian smearing of the reconstructed recoil energy around the
true one, with an energy-dependent width $\sigma_T(T) = \sigma_{T,0}
\sqrt{T/T_\mathrm{th}}$, where $\sigma_{T,0}$ is the energy resolution at the
detection threshold ($T_\mathrm{th}$) listed in Table~\ref{tab:detectors}.  
As for the time resolution, for both  the HPGe PPC and the S1 events in TPCs we assume
Gaussian smearing of the reconstructed time around the true one ---the latter spanning from $t=0$ to $t = t_\mathrm{max}=\infty$---, 
with a constant width $\sigma_t$ given in Table~\ref{tab:detectors}.
For the Cryogenic-CsI detector we find the time resolution to be better described by a
lognormal distribution of $(t^{\mathrm{r}}-t)$ ---so $t_{\rm max}=t^\mathrm{r}$---
centered at  $\mu=2.578$ and with $\sigma=0.7234$, both in  $\ln$ (ns).

For all scenarios considered the true event distribution can be always factorized
as
\begin{equation}
  \frac{d^2N^S}{dt dT} =  \sum_{X=\nu_e,\nu_\mu,\overline{\nu}_\mu} f_{t,\, X}(t)  g_{X}(T)
\end{equation}
where $f_{t,\,X}(t)$ describes the time distribution  (normalized to 1 for convenience)
of the neutrinos with flavour $X$ arriving at the detector.
As described above, neutrinos are produced from pion decay
$\pi^+ \rightarrow \mu^+ + \nu_\mu$,  and subsequent muon decay $\mu^+ \rightarrow
e^+ + \nu_e + \bar{\nu}_\mu$; with both decays happening at rest. Since pions are produced instantaneously when
protons hit the target (as mentioned above, we set that time to be $t=0$) one can obtain the neutrino time distributions as
\begin{align}
  f_{t,\, \nu_\mu}(t) =
  & \int_0^t \mathrm{d}t' \, \frac{\mathrm{d}n_p}{\mathrm{d}t'} (t') \frac{e^{\frac{t'-t}{\tau_\pi}}}{\tau_\pi} \, ,\\
  f_{t,\, \nu_e}(t) = f_{t,\, \bar{\nu}_\mu}(t) = & \int_0^t \mathrm{d}t' \, \frac{\mathrm{d}n_p}{\mathrm{d}t'} (t')
  \frac{e^{\frac{t'-t}{\tau_\mu}} - e^{\frac{t'-t}{\tau_\pi}}}{\tau_\mu - \tau_\pi} \, , 
\end{align}
where $\tau_\pi$ and $\tau_\mu$ are the pion and muon lifetimes, respectively.
$\mathrm{d}n_p/\mathrm{d}t$ is the time profile of the proton beam, which as mentioned above at
J-PARC consists of square pulses of 100 ns separated by 540 ns~\cite{JSNS2:2017gzk}. Explicitly,
\begin{equation}
    \frac{\mathrm{d}n_p}{\mathrm{d}t} = \begin{cases}
        0 & \, \text{if }t<0 \\
        5\,\mathrm{\mu s^{-1}} & \, \text{if }0 < t < 0.1\,\mathrm{\mu s} \\
        0 & \, \text{if }0.1\,\mathrm{\mu s} < t < 0.54\,\mathrm{\mu s} \\
        5\,\mathrm{\mu s^{-1}} & \, \text{if }0.54 < t < 0.64\,\mathrm{\mu s} \\
    \end{cases} \,\, ,
\end{equation}
where we have normalized the time distribution to 1.

The time distributions $f_{t,\,X}(t)$ are the same for all the scenarios considered, 
so that all model-dependence is in the energy-dependent functions $g_{X}(T)$
given by:
\begin{equation}
\!\!\!\!  g_{X}(T) = \varepsilon\, \frac{N_p f_{\nu/p} N_t}{4\pi\ell^2} \int_{E_\nu^\mathrm{min}}^\infty
  \!\!\!\!\mathrm{d}E_\nu \, \phi_X(E_\nu) \frac{\mathrm{d}\sigma_X}{\mathrm{d}T} (E_\nu, T) \, ,
 \label{eq:gt} 
\end{equation}
where $E_\nu^\mathrm{min} = \left.\left(T + \sqrt{T^2 + 2 T M}\right)\right/2$ is
the minimum neutrino energy that can generate a nuclear recoil energy $T$, with $M$ the mass of the
target nucleus. $\phi_X(E_\nu)$ is the normalized neutrino flux of flavour $X$
\begin{align}
\phi_{\nu_\mu} & = \delta\left(E_\nu - \frac{m_\pi^2 - m_\mu^2}{2 m_\pi}\right) \, , \\
\phi_{\nu_e} & = \frac{96}{m_\mu^4} (m_\mu E_\nu^2 - 2 E_\nu^3) \, \theta\left(\frac{m_\mu}{2} - E_\nu\right)\, ,\\
\phi_{\bar{\nu}_\mu} & = \frac{16}{m_\mu^4} (3 m_\mu E_\nu^2 - 4 E_\nu^3) \, \theta\left(\frac{m_\mu}{2} - E_\nu\right) \, .
\end{align}
$m_\mu$ and $m_\pi$ are the pion and muon masses, and $\theta$ is the Heaviside step function,
that cuts the latter two distributions at $E_\nu = m_\mu/2$.

The normalization factors in Eq.~\eqref{eq:gt} include the detection efficiency, $\varepsilon$;
the number of pions per proton, $f_{\pi/p}$; the source-target distance, $\ell$;
the number of target nuclei, $N_t$; and the number of protons on target, $N_p$. The latter is
determined by the beam power, $P$; the proton energy, $E_p$; and the running time, $t_{\rm run}$; as
$N_p = \frac{P}{E_p} \times t_\mathrm{run}$. Unless otherwise stated, a common set of
assumptions apply to these normalization factors, for all detector
configurations considered in this work:
\begin{enumerate}
\item The beam power is assumed to be 1.3\,MW and the proton energy 3\, GeV, as discussed in Sec.~\ref{sec:intro}. 
We assume 3 years of running time and an exposure
  of 3,600\,hr per calendar year due to periods of maintenance at the facility \cite{JPARC_2020, JPARC_2021, JPARC_2022}. Thus
  $$N_p \simeq 3.5 \cdot 10^{22} \left(\frac{P}{1.3\,\mathrm{MW}}\right) \left(\frac{3\,\mathrm{GeV}}{E_p}\right) \left(\frac{t_\mathrm{run}}{\rm yr}\right)\, .$$
\item We assume $f_{\pi/p}=0.44$, as discussed in Sec.~\ref{sec:flux}.
\item The detector distance to the target is set to 20~m. A number of possible detector locations can be envisioned at this distance within the third floor terrace of the MLF \cite{jsns2_2,Ajimura_2015}.

\item The detection efficiency is assumed to be a step function
  at threshold, with a conservative 80\% acceptance.
\end{enumerate}

In the SM, the differential cross section for CE$\nu$NS of $\nu_\alpha$ on a nucleus of mass
$M$ consisting of $Z$ protons and $N$ neutrons reads~\cite{Freedman:1973yd}:
 \begin{equation}
\!\!\!\! \!\frac{d\sigma^\alpha}{dT} \!\!=\!
 \frac{G_F^2}{2\pi} \frac{{\cal Q}_{W,\alpha}^2}{2} |F(Q^2)|^2 M \left[1 - \frac{M T}{2E_\nu^2} -\frac{T}{E_\nu} + \frac{T^2}{2E^2_\nu} \right]\, ,
 \label{eq:xsec-SM}
\end{equation} 
 where $E_\nu$ is the incident neutrino energy, $G_F$ is the Fermi constant, and
\begin{equation}
  {\cal Q}_{W,\alpha} = -N + Z\left[1 - 4 \sin^2 \theta_W(1+\Delta_\alpha)\right]
\label{eq:QWSM}  
\end{equation}
is the weak charge of the target nucleus with $\theta_W$ the Weinberg angle that, for the sake of concreteness, in our SM calculations
has been set to its value at zero momentum transfer $\sin^2\theta_W = 0.23867$ ~\cite{Erler:2017knj}.
In Eq.~\eqref{eq:QWSM} we have introduced the flavour-dependent factor $\Delta\alpha$, which in the SM
is directly related to the effective neutrino charge radius, $\rnu$ ~\cite{Giunti:2014ixa}
defined as 
\begin{equation}
\label{eq:rnu}
\rnu \equiv -6 \frac{dF^{\rm em}_\nu(Q^2)}{dQ^2} \bigg |_{Q^2 = 0} \, ,
\end{equation}
where $F^{\rm em}_\nu$ is the electromagnetic form factor of the neutrino. The
inclusion of this form factor affects the scattering of neutrinos with
other charged particles and effectively induces a flavour-dependent
shift~\cite{Degrassi:1989ip, Vogel:1989iv, Kouzakov:2017hbc}, 
\begin{equation}
\label{eq:thetaW-rnu}
\Delta_\alpha= \frac{1}{3}m_W^2\langle r_{\nu_\alpha}^2\rangle 
\end{equation}
where $m_W$ is the mass of the $W$ boson. The value of the neutrino charge radius in the SM~\cite{Bernabeu:2000hf, 
Bernabeu:2002nw, Bernabeu:2002pd} is
\begin{equation}
\langle r_{\nu_\alpha}^2\rangle 
= \frac{- G_F}{2\sqrt{2} \pi^2}
\left[ 3 - 2 \ln \left( \frac{m_{\ell_\alpha}^2}{m_W^2}\right) \right] \, ,
\label{eq:rnu-SM}
\end{equation}
with $m_{\ell_\alpha}$ the mass of the charged lepton of
flavour $\alpha$.
In Eq.~\eqref{eq:xsec-SM}, $F(Q^2)$ is the weak form factor of the nucleus evaluated at the squared
momentum transfer of the process, $Q^2 = 2 M T$, which we parametrize as~\cite{Helm:1956zz}
\begin{equation}
  F(Q^2) = 3 \frac{j_1(Q R_0)}{Q R_0} e^{-\frac{Q^2 s^2}{2}}
\label{eq:weakff}  
\end{equation}
with $j_1$ being the spherical Bessel function, $R^2_0 \equiv \frac{5}{3}(R_W^2 - 3 s^2)$, 
$R_W$ the weak radius of the nucleus
\begin{equation}
  R_W^2\equiv-6 \frac{dF(Q^2)}{dQ^2} \bigg |_{Q^2 = 0} \, ,
\label{eq:rwdef}
\end{equation}  
and $s = 0.9\,\mathrm{fm}$ the neutron skin.  As CE$\nu$NS happens at low momentum transfers, 
only the low-Q$^2$ values of the form factor are phenomenologically relevant, and other parametrizations of $F(Q^2)$ with the
same weak radius would lead to very similar results.

$R_W$ can be expressed in terms of the nuclear charge radius ($R_{\rm ch}$),
the point-proton and neutron distribution radius ($R^{\rm pt}_p$ and $R^{\rm pt}_n$), and the proton and neutron
square charge radius ($r_{\mathrm{ch},\,n}$  and $r_{\mathrm{ch},\,p}$) as~\cite{Coloma:2020nhf}
\begin{equation}
 \!\!\!R_W^2 \!=\! R_\mathrm{ch}^2 \!+\! \frac{N}{Q_W} \left[\!(R^{\rm pt}_n)^2\! -\! (R^{\rm pt} _p)^2 \!+\! \frac{Z^2\!-\!N^2}{Z N} r_{\mathrm{ch},\,n}^2\right]
\label{eq:rw}
\end{equation}
with $(R^{\rm pt}_p)^2 \equiv R_\mathrm{ch}^2 - r_{\mathrm{ch},\,p}^2- \frac{N}{Z}r_{\mathrm{ch},\,n}^2$.
In our calculations we use as inputs the tabulated nuclear charge
radii from Ref.~\cite{Angeli:2013epw}, together with the Particle Data
Group (PDG)~\cite{ParticleDataGroup:2024cfk} values for the squared proton
and neutron charge radii: $r_{\mathrm{ch},\,p}^2 =
0.707\,\mathrm{fm^2}$ and $r_{\mathrm{ch},\,n}^2 =
-0.1161\,\mathrm{fm^2}$. For the sake of concreteness, in our simulations we generate the expected number of events
by setting the point neutron radius
to be the best fit value of the fit in Ref.~\cite{Trzcinska:2001sy}, 
$R_n^{\rm pt,true}=R_p^{\rm pt}-0.04+1.01 N-Z/A$.

The CE$\nu$NS cross section depends on the target nuclear isotope. For Xe, Ge, and Ar,
we use their natural isotope abundances as taken from Ref.~\cite{elementIsotopes}.
For the CsI detector we assume a 50\% number abundance of Cs and 50\%
of I. For convenience, the values of $R_{\rm ch}$
(and hence $R^{\rm pt}_p$) as well as the assumed true value of $R^{\rm pt,true}_n$ used in
our calculations are summarized in Tab.~\ref{tab:nuclei}. 
\begin{table}
\centering  
\begin{tabular}{| c | cccccc |}
\hline
Target & $Z$ & $N$ &  \% & $R_{ch}$ (fm) & $R^{\rm pt} _p$ (fm)& $R^{\rm pt, true}_n$ (fm)
\\ \hline  
Xe & 54 &  78 & 26.9 & 4.786 &4.729 & 4.873\\
&    &  75 & 26.4 & 4.778 &4.720 & 4.844 \\
&    &  77 & 21.2 & 4.780 &4.724 & 4.861 \\
&    &  80 & 10.4 & 4.780 &4.734 & 4.890 \\
&    &  82 & 8.9  & 4.796 &4.741 & 4.909 \\
&    &  76 & 4.1  & 4.782 &4.725 & 4.885 \\
&    &  74 & 1.9  & 4.777 &4.720 & 4.837 
\\\hline
Ar & 18 & 22 &100 & 2.427 & 2.308 & 2.369\\\hline
Ge & 32 &  42 & 36.7 & 4.074 & 4.005 & 4.102\\
  &   &  40  & 27.3 & 4.058 & 3.988 & 4.060\\
  &   &  38  & 20.4 & 4.041 & 3.970 & 4.017\\
  &   &  44  & 7.8  & 4.081 & 4.013 & 4.133\\
  &   &  41  & 7.8  & 4.063 & 3.994 & 4.078  
\\\hline
CsI &  55  & 78 & 50 & 4.750 & 4.692 & 4.819\\
    &  53  & 74 & 50 & 4.804 & 4.747 & 4.882
\\\hline
\end{tabular}
\caption{Main properties of the nuclei/molecule for the different targets
  considered in this work. The different columns indicate the isotope
  considered and their abundance together with the number of protons and neutrons,
  the nucleus charge radius, the point proton distribution radius, and the true point neutron
  distribution radius assumed in the simulations.} 
\label{tab:nuclei}
\end{table}

In what respects the expected backgrounds, they can be divided into
three classes (see Sec.~\ref{sec:bckg_sources} for more details): 
(\textit{i}) steady-state backgrounds (SS bck); (\textit{ii})
beam-related backgrounds (BR bck), produced by neutrons escaping the
target monolith and reaching the detector; and (\textit{iii}) neutrino-induced
neutrons (NiN bck).  For each of the detectors considered the time and
energy dependence of these three backgrounds have been estimated 
as discussed in Sec.~\ref{sec:detectors}. For illustration we list in table~\ref{tab:detectors} the
total number of expected CE$\nu$NS events and of the three considered
backgrounds per run year (including the 80\% detection efficiency).
It is important to notice that, while the steady-state background can
be in principle sizable, it can be efficiently measured using beam-off
data and therefore can be subtracted leaving only the residual
statistical uncertainty associated to the subtraction process. 
\begin{figure*}[ht!]
\includegraphics[width=0.9\textwidth]{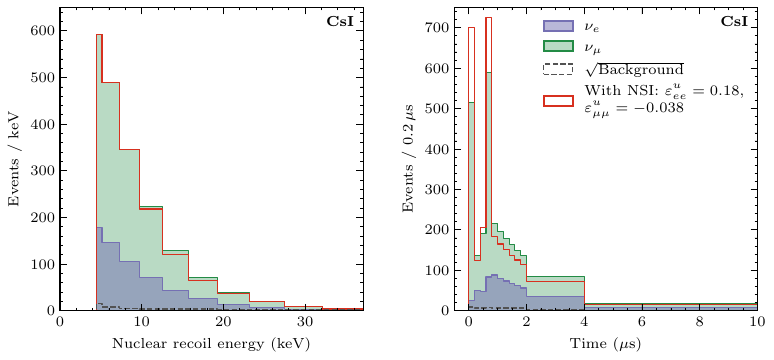}
\caption{Predicted
event distribution as a function of the reconstructed recoil energy,
$T^{\rm r}$, and the reconstructed time $t^{\rm r}$ in the Cryogenic-CsI detector
(with the steady-state background subtracted) for three years of run time.
The distributions are shown for the SM and for an example of NSI scenario (see below for definitions).
The binning shown is the one employed in the analysis for this detector and it has been
chosen to ensure enough events when binning in both variables.}
\label{fig:speccsi}  
\end{figure*}

As described in Sec.~\ref{sec:flux}, the pulsed nature of the J-PARC proton beam
provides a unique handle to discriminate BSM flavour-dependent scenarios. To 
illustrate this, we plot in Fig.~\ref{fig:speccsi} the predicted
distribution of events as a function of the reconstructed recoil energy,
$T^{\rm r}$; and the reconstructed time, $t^{\rm r}$; in the Cryogenic-CsI detector
(with the steady-state background subtracted) for three years of run time.
The figure clearly displays how  the relative weight of $\nu_e$-induced and
$\nu_\mu$-induced event substantially changes with the time of the event.
To this end, the distribution is shown for the SM as well as for an NSI scenario
(see below for definitions) which leads to almost the same total number of events.
As seen in the left panel, the recoil energy distribution does not allow to discriminate
this model from the SM. On the contrary, as seen on the right panel,  with the time information
this scenario can be tested with high confidence.

Altogether, in order to determine the sensitivity to a  model
characterized by a set of parameters $\left\{ \varepsilon
\right\}$, we build a bi-dimensional binned $\chi^2$. 
Systematic uncertainties are implemented using the pull method.
We consider three sources of systematic uncertainties which we assume to be fully correlated
among the bins: 10\% uncertainty for the  flux normalization
---see the discussion in Sec.~\ref{sec:flux}--- , an achievable 5\% uncertainty on  
the QF \cite{csina, Lewis:2021cjv}---equivalent to an energy-scale uncertainty---
(pull $\xi_{QF}$ with  $\sigma_{QF}=0.05$), 
and 5\% uncertainty on the normalization of the irreducible  beam-related
and  neutrino-induced-neutron backgrounds (pull $\xi_{B}$ with
$\sigma_{B}=0.05$). We have verified that including two different pulls for these
two backgrounds leads to very similar results.

With this, for HPGe PPC and Cryogenic-CsI detectors and for the S1 TPC events,
we construct:
\begin{equation}
\!\!\!\!\chi^2\big( \{\varepsilon\}\big) = {\rm min}_{\xi}
\left[
  \chi^2\big(\{\varepsilon\},\xi\big) +\!\!\!\!\!\!\!\!
  \sum_{p=F,QF,B}
\left(\frac{1-\xi_{p}}{\sigma_{p}}\right)^2 \right]
\,  ,
\label{eq:chi2}
\end{equation} 
where
\begin{eqnarray}
\chi^2 (\left\{ \varepsilon \right\},\xi ) = \sum_{ij} 2 \Big[&& N_{ij} (\{
  \varepsilon \},\xi ) - \overline{ N}_{ij}  \\\nonumber
&&+ \overline{N}_{ij} \ln \left(\frac{\overline{N}_{ij}}{N_{ij} (\{ \varepsilon\},\xi ) }
  \right) \Big] .
\end{eqnarray}
$N_{ij}(\{ \varepsilon\},\xi )$ are the predicted number of events 
with reconstructed time
$\in [t_i^\mathrm{r}, t_{i+1}^\mathrm{r}]$ and reconstructed nuclear recoil energy $\in [T_j^\mathrm{r}, T_{j+1}^\mathrm{r}]$ for the model to be tested, 
\begin{eqnarray}
  N_{ij} (\{ \varepsilon\},\xi )   &=&
  \xi_F \int_{t_i^\mathrm{r}}^{t_{i+1}^\mathrm{r}} \!\!\!\!\!d t^\mathrm{r}
  \int_{\xi_{QF} T_j^\mathrm{r}}^{\xi_{QF} T_{j+1}^\mathrm{r}} \!\!\!\!\!dT^\mathrm{r}
  \frac{d^2 N^S}{dt^{\rm r} dT^{\rm r}} (\{ \varepsilon\})\nonumber\\ &&\!\!\!\!\!\!+ N^{\rm bck}_{{\rm SS},ij}
  +\xi_B (N^{\rm bck}_{{\rm BR},ij}+N^{\rm bck}_{{\rm NiN} ,ij})\;.
\label{eq:nij}  
\end{eqnarray}
$\overline{N}_{ij}$ stands for the {\sl data} event rates which we assume to be the rates expected
in that bin from the combination of signal and background in the SM with $\xi_i=1$, and
for the assumed true values of the SM parameters. We notice that in Eq.~\eqref{eq:nij} we do not
include any systematic background pull for the steady-state background since, as mentioned above,
we assume that this background  can be independently measured using beam-off data, and therefore
subtracted, so we account only for its effect on the statistical uncertainty.

In addition, for high pressure TPC detectors  we add the corresponding $\chi^2$
for the NoS1 events which is only binned in $T^{\rm r}$. 

\subsection{Weak mixing angle and neutrino charge radii }
\label{sec:thetaW}
The weak mixing angle is a fundamental parameter in the SM whose scale
dependence in a given renormalization scheme is predicted within the
model, with a precision of $2\times 10^{-5}$ ~\cite{Erler:2017knj}, to
become approximately constant at scales below ${\cal O}(0.2)$ GeV.
Experimentally, it has been most precisely measured at the electroweak
scale in collider experiments ~\cite{ParticleDataGroup:2024cfk},
with ${\cal O} (10^{-3})$ or better accuracy. This sets the
target precision for low-energy experiments to provide a test of the
model.  Below GeV scales, $\sin^2\theta_W$ has been determined at
$\sim 0.2$ GeV in measurements of the parity-violating asymmetry in
Moller scattering~\cite{Anthony:2005pm} and electron-proton scattering
~\cite{Qweak:2018tjf} with $\sim 0.5$ \% precision, and at $\sim 2.4$
MeV in measurements of parity violation in Cs
atoms~\cite{Wood:1997zq,Dzuba:2012kx} with $\sim 0.8$ \% precision.
CE$\nu$NS provides also a low-energy measurement of $\sin^2\theta_W$, 
with the most precise determination so far provided by  the
COHERENT measurement of CE$\nu$NS on CsI,
$\sin^2\theta_W= 0.220^{+0.028}_{-0.026}$ (1$\sigma$)~\cite{COHERENT:2021xmm}.

\begin{table*}
\renewcommand{\arraystretch}{1.4}
\begin{tabular}{|c|cccc|}
    \hline
  & CsI & Ge & Xe & Ar
\\  
\hline 
$\sin^2 \theta_W$ &
$ 0.239^{+ 0.033}_{-0.026}$ & $ 0.239^{+ 0.031}_{-0.025}$ &
$ 0.239^{+ 0.033}_{-0.027}$ &$0.239^{+ 0.037}_{-0.033}$ \\
\hline
$\langle r^2_{\nu_e}\rangle\;(10^{-32})$ cm$^2$ &
$[ -54.,-48.3]\oplus[  -3.,  2.1]$&
$[ -42.,-36.6]\oplus[ -15.,-10.1]$&
 $[ -56.,-46.3]\oplus[  -5.,  4.7]$&
 $[ -36.,-15.7]$\\
$\langle r^2_{\nu_\mu}\rangle\;(10^{-32})$ cm$^2$ &
 $[ -54.,-49.1]\oplus[  -3.,  2.1]$&
 $[ -41.,-37.8]\oplus[ -14.,-10.6]$&
 $[ -55.,-49.0]\oplus[  -3.,  3.4]$&
 $[ -35.,-31.2]\oplus[ -20.,-16.8]$
\\
$\frac{\langle r^2_{\nu_e}\rangle}{\langle r^2_{\nu_\mu}\rangle_{\rm SM}}=
  \frac{\langle r^2_{\nu_e}\rangle}{\langle r^2_{\nu_\mu}\rangle_{\rm SM}}$
    & $1^{+2.9}_{-3.5}$ & $1^{+3.0}_{-3.8}$& $1^{+3.2}_{-4.1}$& $1^{+4.0}_{-4.6}$
 \\   
 \hline
 $R^{\rm pt}_n$ (fm) &
 $4.85^{+0.37}_{-0.39}$ & $4.07^{+0.72}_{-0.83}$ &  $4.86^{+0.29}_{-0.35}$ & $<4.24$
\\\hline 
$\mu_{\nu_e}$ ($10^{-10} \mu_{\rm B}$) &
$< 16$ &  $< 9.4$ & $< 11$ & $<14$\\
$\mu_{\nu_\mu}$ ($10^{-10} \mu_{\rm B}$) &
$< 12$ & $< 7.7$ & $<8.8$ &  $< 11$\\\hline
$\left. g_Z'^{B-L}\right|_{M_{Z'}\lesssim 5\,{\rm MeV}}$
& $<1.4\times10^{-5}$ &$<6.3\times 10^{-6}$ & $<7.1\times 10^{-6}$ & $<8.8\times 10^{-6}$ \\
$\left.\frac{g_{Z'}^{B-L}}{M_{Z'}/GeV}\right|_{M_{Z'}\gtrsim 50\,{\rm MeV}}$ &
$<4.9 \times 10^{-4}$ &$<3.9\times 10^{-4}$ & $<3.5\times 10^{-4}$ & $<5.3\times 10^{-4}$ \\
\hline
\end{tabular}
\caption{Allowed ranges at 90\% C.L. for the weak mixing
  angle (given as best fit $\pm 1.64 \sigma$), neutrino charge radii for two flavour
  projections  (and after marginalizing over  the other), the point-neutron
  radius,   the $\nu_{e,\mu}$ magnetic moments, and the model with a light vector mediator coupled to $B-L$
  current  
  (90\% C.L. upper bound)} 
\label{tab:macrotable}
\end{table*}

Our results on the expected sensitivity for this parameter are shown in
Table~\ref{tab:macrotable} for the different detectors under
consideration.
As seen in the table, any of the experiments considered here leads only 
to a mild improvement on the determination of the weak mixing angle in
CE$\nu$NS with respect to the current COHERENT result, all being still far from the
precision of atomic parity violation experiments. 
This is so, because, as seen in Eq.~\eqref{eq:xsec-SM}, the weak mixing angle enters
the CE$\nu$NS cross section through the coupling with protons $Z$ in the weak charge
$Q_W$, leading only to an overall normalization correction to the total number of expected events.
Consequently, the sensitivity  to this parameter for all detectors considered
is comparable because it is mostly limited by the assumed 10\% flux normalization uncertainty.
To this end, we have explicitly verified that removing the QF uncertainty and/or
the time information has no effect on these results. The ultimate 1$\sigma$ statistics-limited sensitivity that could be 
achieved if there were no flux uncertainty is (1.3, 2.9, 1.2, 5.6) \%  at 1$\sigma$ for the
(CsI, Ge, Xe, Ar)-based detectors. 

As mentioned above, the determination of the weak mixing angle is tightly related to the
sensitivity to the effective flavour-dependent neutrino  charge radius whose value in the  
SM~\eqref{eq:rnu-SM} is
\begin{equation}
  \langle r_{\nu_\alpha}^2 \rangle
  =-\left(\begin{array}{c}
    \!\!0.83\\[-0.1cm]\!\!0.48\\[-0.1cm]\!\!0.3\end{array}\!\!\right) \times 10^{-32}\,{\rm cm}^2\;{\rm for}\;
    \left(\!\!\begin{array}{c}\nu_e\\[-0.1cm]\nu_\mu\\[-0.1cm]\nu_\tau \end{array}\!\!\right)
    \label{eq:rnusm}
    \end{equation}
making the observation of this effect extremely challenging (see Ref.~\cite{Giunti:2024gec}
for a  recent review and compilation of bounds).

\begin{figure*}[ht!]
\includegraphics[width=0.45\textwidth]{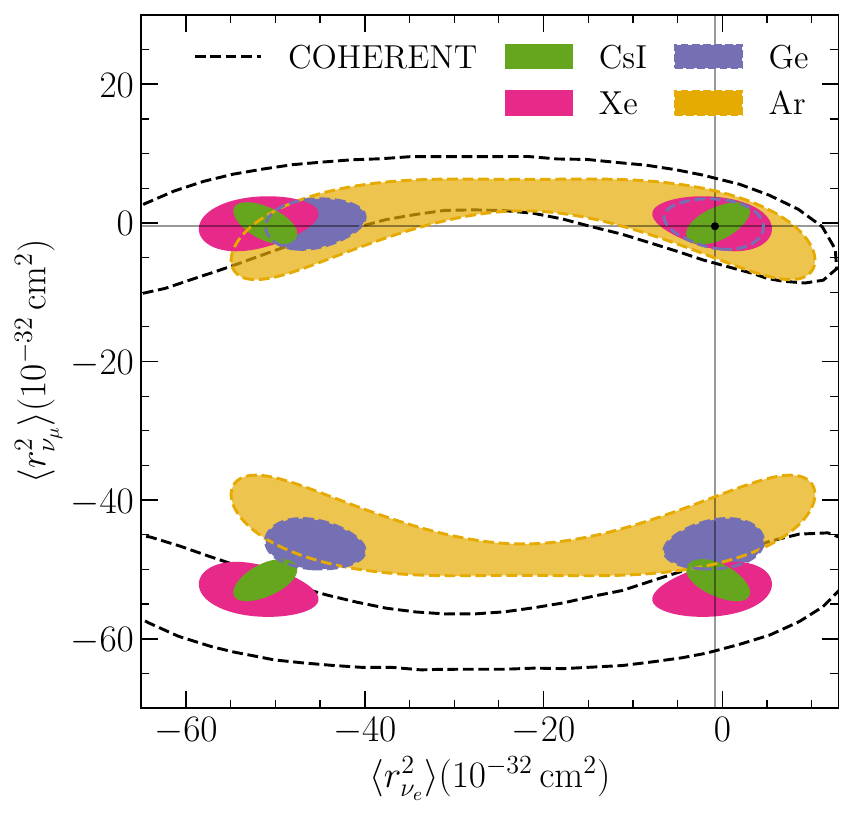}
\includegraphics[width=0.45\textwidth]{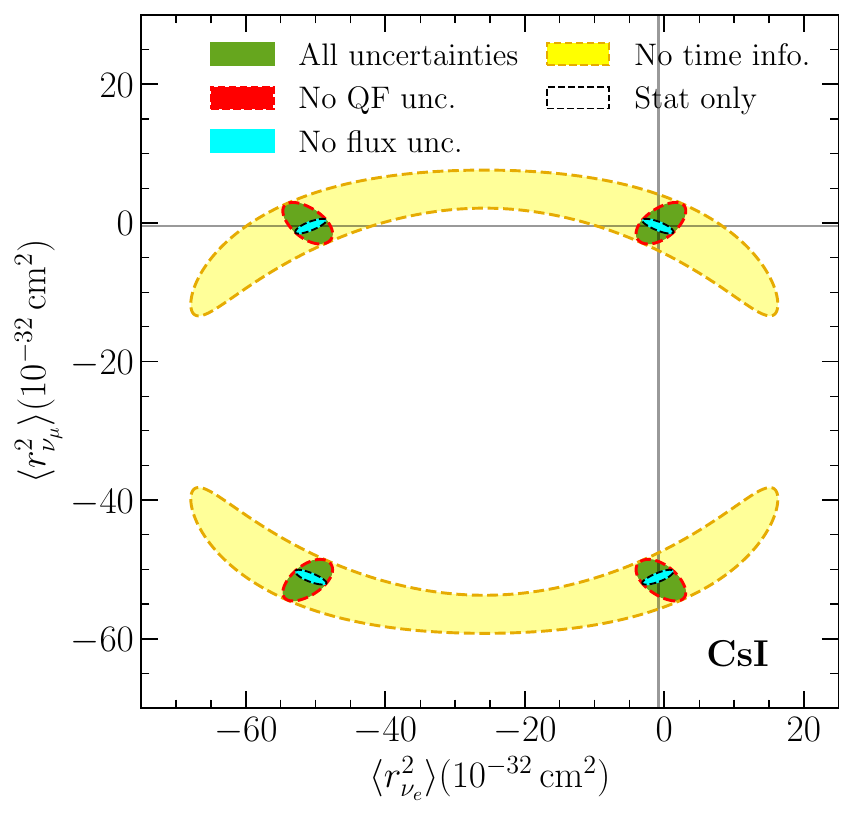}
\caption{Expected allowed regions in the
  $(\langle r_{\nu_e}^2 \rangle,\langle r_{\nu_\mu}^2 \rangle)$
   plane at the 90\%
   confidence level (C.L.) for two degrees of freedom (d.o.f.)  ($\Delta\chi^2=4.61$)
   . {\bf Left:} The
  different regions correspond to the expected results for the
  different detectors listed in Table~\ref{tab:detectors}, as
  indicated by the legend. In all cases, the simulated data has been
  generated for the SM, and the results are then fitted assuming arbitrary values of the
  charge radii and including the nominal uncertainties and time resolution. For comparison
  we plot as the vertical and horizontal lines the SM predictions (Eq.~\eqref{eq:rnusm}), and as dashed the current
    90\%CL bounds from the analysis of COHERENT CsI and Ar data from
    Ref.~\cite{DeRomeri:2022twg} (see also
    Refs.~\cite{AtzoriCorona:2022qrf, AtzoriCorona:2024rtv}).
  {\bf Right:}
  The allowed regions for the Cryogenic-CsI detector under different assumptions
  for the systematic uncertainties as labeled, see text for details. }
\label{fig:rsq}
\end{figure*}

In BSM scenarios, the charge radii may receive additional contributions
and, therefore, a measurement of their value well above the SM
expectation would be a clear signal of BSM. In addition, in some BSM
scenarios flavour transition charge radii $\langle
r_{\alpha\neq\beta}^2 \rangle$ can also be generated~\cite{Cadeddu:2019eta,Kouzakov:2017hbc}.  For concreteness, we
study the expected sensitivity for scenarios generating flavour-diagonal 
charge radii only.

Our results on the expected sensitivity for the flavour-diagonal
charge radius  are shown in Fig.~\ref{fig:rsq} and Table~\ref{tab:macrotable}.
In the left panel of Fig.~\ref{fig:rsq},
we plot the 90\% C.L. allowed regions in the plane
$(\langle r_{\nu_e}^2 \rangle,\langle r_{\nu_\mu}^2 \rangle)$
for the  four detectors under consideration. We list the corresponding
90\% C.L. allowed ranges for each charge radius (after marginalizing over the other)
in Table~\ref{tab:macrotable}.
As seen in the figure, for most detectors there are four distinct
regions of which only one is centered at the SM predictions.
The other three correspond to quasi-degenerate scenarios
for which the charge radius is large enough to flip the sign of
the SM $Q_{W,e}$ and/or $Q_{W,\mu}$ (CE$\nu$NS is sensitive to the \emph{squared} weak charge). This happens when
\begin{equation}
  \langle r_{\nu_{e,\mu}}^2 \rangle\simeq -\frac{3}{m_W^2}\left( \frac{N-Z}{2Z\sin^2\theta_W}+2\right)\;.
\label{eq:rsqdeg}  
\end{equation}
For the Ar TPC, the limited time resolution and large backgrounds imply that the interaction
of different flavours cannot be experimentally separated. Hence, data is not very sensitive to $Q_{W,e}^2$
and $Q_{W, \mu}^2$ separately, but to a combination instead (below we discuss this in more detail)
and there is a continuous degenerate region where the combination has the same value as in the SM. 
The figure also illustrates the complementarity of the different targets and how combination of data 
taken with them (i.e., different N/Z) can help to break the 2$\times$2 fold degeneracy 
and isolate a single region.

For comparison, in the curve labeled COHERENT we show the current
  constraints from the observation of CE$\nu$NS at the spallation
  source experiment COHERENT, as obtained in
  Ref.~\cite{DeRomeri:2022twg} by the combined analysis of results
  with CsI~\cite{COHERENT:2021xmm} and Ar~\cite{COHERENT:2020iec}
  targets.  Similar results were obtained in
  Refs.~\cite{AtzoriCorona:2022qrf, AtzoriCorona:2024rtv} with a
  momentum-independent form factor (the assumption in our
  analysis). As seen from the figure, for all detectors under
  consideration JPARC is expected to improve over current bounds from
  the COHERENT CsI+Ar combination. The improvement is substantial for
  detectors with better time resolution (Cryogenic-CsI and HPGe), for
  which flavour discrimination is better; or detectors with moderate
  time resolution but negligible backgrounds (Xe TPC).

The role of the different sources of uncertainties is further illustrated in the right panel
of Fig.~\ref{fig:rsq}, where we show the
effect of the different uncertainties and time information in the final sensitivity
(here for the Cryogenic-CsI detector as illustration, for other detectors the results are 
similar). The different regions are obtained removing one effect at a time. As seen in the figure, 
comparing the red contour and the green region, removing the QF uncertainty leads to very little effect,
as the impact of the charge radius is energy-independent.
Conversely, the results obtained removing
only the flux uncertainty (cyan region) and removing all systematics (black dash contour)
are practically equivalent.
We conclude that the overall flux normalization is the dominant systematic error in this scenario. 
We have also verified that the systematic uncertainty on the intrinsic backgrounds
has a very marginal effect on the results. This is expected given the smallness of
the backgrounds (see Fig.~\ref{fig:speccsi}). This result holds for all the physics scenarios
we have studied.

We also illustrate the relevance of the time information
by showing the allowed region (yellow) if we only bin in energy. In this case, the
region becomes almost a ring. This is so because without time information the dominant effect is a
global correction of the event rates in all energy bins. Requiring that the total number
of events is compatible with the SM
expectation, it is straightforward to show that (neglecting the small SM $\nu$ charge radius),
the allowed confidence
regions in the plane $(\langle r_{\nu_e}^2 \rangle,\langle r_{\nu_\mu}^2 \rangle)$ verify
the equation of a circle
\begin{equation}
\label{eq:ellipse}
\left[ R + \langle r^2_{\nu_{e}}\rangle\right]^2 + 
2\left[ R + \langle r^2_{\nu_{\mu}}\rangle\right]^2 = 3 R^2
\end{equation}
where $R \equiv \frac{3}{m_W^2}\frac{N-Z(1-4\sin^2\theta_W)}{4 Z \sin^2\theta_W}$.
As seen in the figure, this full degeneracy is not exact. This is so because, as discussed in
detail in Ref.~\cite{Baxter:2019mcx},
a detector with enough energy resolution,
large statistics, and no saturation, has some residual flavour discrimination due to the
different flavour composition of the beam  above and below the maximum recoil allowed
for the prompt $\nu_\mu$  component. Nevertheless, the figure clearly illustrates the limited reach
of this effect compared to the flavour-discriminating power of the timing information.

Finally, we note that the $2\times2$ fold degeneracy in  Eq.~\eqref{eq:rsqdeg}  
is absent in models in which the correction to the SM is flavour-independent.
In this case, only values around the SM ones are allowed. Then, our results 
in Table~\ref{tab:macrotable} show that the attainable precision at J-PARC MLF is comparable 
with the SM prediction. Indeed, the statistical uncertainty is small enough that without 
flux normalization uncertainties the SM value could be distinguished from 0. However, 
the flux uncertainty dominates the final result, making the overall uncertainty 3--4 
times larger than the difference between the SM value and 0.

\subsection{Neutron nuclear distribution radius}

\begin{figure*}[ht!]
\includegraphics[width=0.55\textwidth]{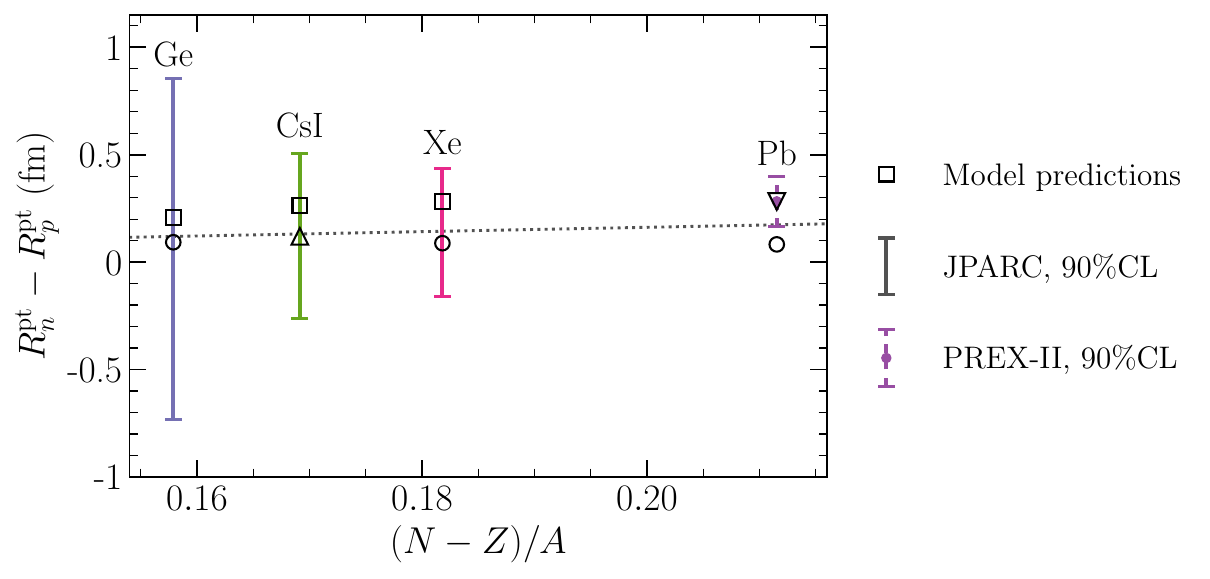}
\includegraphics[width=0.33\textwidth]{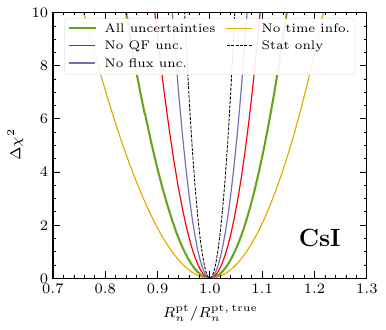}
\caption{{\bf Left:}
  Compilation of our results for the expected sensitivity
  of the neutron skin thickness $R^{\rm pt}_n-R^{\rm pt}_p$ with 90\% C.L., one d.o.f, ($\pm 1.64\sigma$) 
  for the different detectors as labeled in the figure.
  For comparison we show the result for $^{208}$Pb derived from the PREX measurement with the analysis in Ref.~\cite{Horowitz:2012tj}
 and the best-fit to the indirect determination from antiprotonic
 atom x-ray data for a variety of nuclei (dashed line), taken from Ref.~\cite{Trzcinska:2001sy}.
 The hollow markers represent the range of predictions from different nuclear models.
 {\bf Right:} $\Delta\chi^2$ as a function of $R_n^{\rm pt}({\rm CsI})/R^{\rm pt,true}({\rm CsI})$
 under different assumptions for the systematic uncertainties as labeled,
 see text for details. }
\label{fig:rn}
\end{figure*}
Since CE$\nu$NS is sensitive to the weak form factor,
dominated by the coupling to neutrons, it can
directly probe the distribution of neutrons in the nucleus.  This 
provides complementary information to proton densities accessible with
elastic electron scattering~\cite{Donnelly:1984rg,Angeli:2013epw}.  At
present the most precise direct measurement of a neutron distribution comes
from parity-violating electron scattering in
$^{208}$Pb~\cite{Abrahamyan:2012gp, PREX:2021umo}, also sensitive to the weak form
factor.  Alternative measurements rely on
nuclear~\cite{GarciaRecio:1991wk,Suzuki:1995yc,Clark:2002se,Trzcinska:2001sy,Lapoux:2016exf}
or electromagnetic~\cite{Tarbert:2013jze} reactions which probe both
neutron and proton distributions, but they lean on model-dependent
analyses (with uncertainties that are difficult to quantify).
Likewise, atomic parity violation experiments are also sensitive to
the nuclear neutron distribution, but they are subject to
model-dependent uncertainties from atomic many-body
calculations.

Given the low momentum transfers involved in CE$\nu$NS, the form
factors can be characterized using the first moment of the weak charge
distribution in $Q^2$, $R_W$ in Eq.~\eqref{eq:rw}), or what is the
same, in terms of the point-proton $R^{\rm pt}_p$ and point-neutron
distribution radii $R^{\rm pt}_n$. In the last years, different
phenomenological analysis of the data from CE$\nu$NS observations have
provided determinations of the neutron distribution radius on CsI
\cite{Cadeddu:2017etk,Cadeddu:2018dux,Cadeddu:2019eta,Papoulias:2019lfi,Khan:2019cvi,Huang:2019ene,Coloma:2020nhf,AtzoriCorona:2023ktl}, Ar~\cite{Miranda:2020tif,Cadeddu:2020lky}, and
Ge~\cite{AtzoriCorona:2025xgj} with consistent results such as
\begin{eqnarray}  
  R_n({\rm CsI})&=&5.47^{+0.63}_{-0.72}\;{\rm fm}\;  (90\%\,{\rm C.L.}) \;
   { \text{\cite{AtzoriCorona:2023ktl}}}
    \nonumber
    \\
    R_n({\rm Ar})&\leq& 6.2 \;{\rm fm} \;
  (2\sigma\,{\rm C.L.})\;\text{\cite{Cadeddu:2020lky}}
    \\
     R_n({\rm Ge})&=&6.5\pm 3\;{\rm  fm} \; (2\sigma \,{\rm C.L.})\;\text{\cite{AtzoriCorona:2025xgj}}
\nonumber
\end{eqnarray}

Our results on the expected sensitivity for the
point-neutron radius are shown in Fig.~\ref{fig:rn} and
Table~\ref{tab:macrotable}.  In the table we show the expectations for the
95\% C.L. allowed range for each of the considered targets with the
central value corresponding to the true value of the weighted averaged
over the considered isotopes (see table~\ref{tab:nuclei}). These predictions
clearly represent a substantial improvement over our current knowledge. On
that end, we plot in the left panel in Fig.~\ref{fig:rn} our results in
the form of the expected sensitivity of the neutron skin thickness
$R^{\rm pt}_n-R^{\rm pt}_p$ as a function of $(N-Z)/(N+Z)$.  For
comparison, Fig.~\ref{fig:rn} shows the neutron skin thickness for
$^{208}$Pb, derived from the analysis in Ref.~\cite{Horowitz:2012tj}
of the results of the Lead Radius Experiment (PREX)
experiment~\cite{Abrahamyan:2012gp, PREX:2021umo} on parity-violation in electron
scattering which, as mentioned above contains certain model
dependence.  In this figure we also show the best-fit for the neutron
skin obtained from its indirect determination from antiprotonic X-ray
data for a variety of nuclei (dashed line), taken from
Ref.~\cite{Trzcinska:2001sy}.  Finally, the hollow markers show the
predictions for a variety of nuclear models
~\cite{Co:2020gwl,Hoferichter:2018acd,Hoferichter:2020osn}.

The results also show that, unlike for the
Weinberg angle, the attainable precision for the point-neutron radius is worse for
Ge (and much worse for Ar) than for Xe or CsI. 
This is expected on the basis of statistics, because  the signal statistics grows quadratically
with the number of neutrons in the target nucleus. Besides that, the lighter the nucleus
the smaller the values of $R^{\rm pt}_{p,n}$ (or, equivalently, $R_W$), and the weaker
the dependence of the form factor with $Q^2$,  reducing the energy dependence of the
effect and making it more difficult to disentangle from the overall normalization uncertainty.
To illustrate the relevance of the different systematic uncertainties on the attainable
sensitivity we plot in the right panel (Figure~\ref{fig:rn}) the dependence of $\Delta\chi^2$ on $R_n^{\rm pt}(\mathrm{CsI})$
for different assumptions on the systematic uncertainties. First we notice that, while
the flux normalization keeps being the main systematic uncertainty, the assumed precision on the
QF plays also an important role. This is so because the effect of $R_n^{\rm pt}$ is to suppress 
the number of events at large recoil energies as coherence gets lost. The energies at which 
this suppression happens are directly related with the value of $R_n^{\rm pt}$, and a QF uncertainty is
equivalent to an uncertainty in the reconstructed nuclear recoil energy, i.e., an uncertainty
in the reconstructed $R_n^{\rm pt}$. Finally we also observe that despite the effect being 
flavour-independent, there is a certain loss of sensitivity if the time information is not included.
This is due to the different energy dependence of the events from the prompt and delayed
neutrinos. Thus, timing information aids in establishing  the $T$-dependence of the effect.

Overall, we find that statistical and systematic effects are comparable in this measurement. 
As an illustration, if the CsI detector mass were increased up to 1 tonne (assuming that
backgrounds scale up proportionally), we find that the 90\%~C.L. uncertainty on the point neutron radius would 
decrease by a factor $\sim 4$.

\subsection{Non-standard neutrino interactions}
\label{sec:nsi}
\begin{figure}[ht!]
\includegraphics[width=0.47\textwidth]{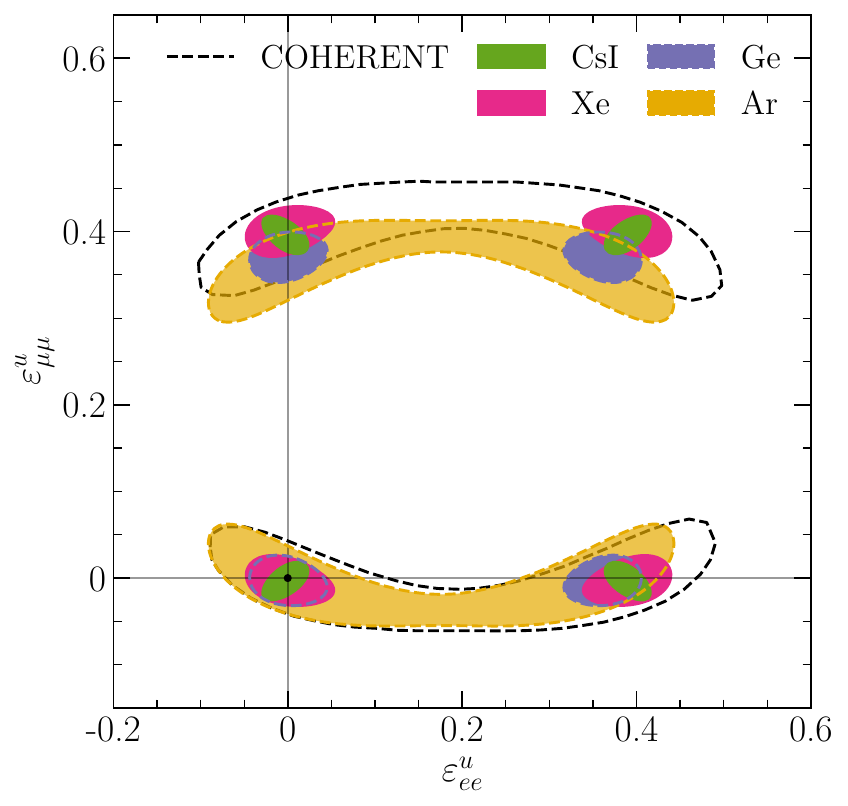}
\caption{Expected allowed regions in the
  $(\epsilon_{ee}^{u,V} ,\epsilon_{\mu\mu}^{u,V})$
  plane at the 90\%
  confidence level (C.L.) for two d.o.f. ($\Delta\chi^2=4.61$).
  The  different regions correspond to the expected results for the
  different detectors listed in Table~\ref{tab:detectors}, as
  indicated by the legend.
  For simplicity, the rest of the NSI parameters not shown in the figure have been
  assumed to be zero.
  In all cases, the simulated data has been
  generated for the SM, and the results are then fitted assuming arbitrary values of the
  two NSI coefficients and including the nominal uncertainties and time resolution. The
    dashed contour shows the current 90\% CL bounds from the analysis
    of COHERENT CsI and Ar data from Ref.~\cite{DeRomeri:2022twg}.
}
\label{fig:NSI}
\end{figure}

From a completely model-independent approach, a useful parametrization
of the possible BSM effects at low energies is through the
addition of higher-dimensional operators to the SM Lagrangian. 
At dimension 6, the allowed
set of operators includes four-fermion operators affecting neutrino
production, propagation, and detection processes. These are the
so-called non-standard neutrino interactions (NSI).
For example, the effective Lagrangian
\begin{equation}
\label{eq:nsi-nc}
\!\!\mathcal{L}^\text{NC}_\text{NSI}\! =\! -\!\! \sum_{f,\alpha,\beta}\!\!\! 2\sqrt{2} \, G_F \, \Eps_{\alpha \beta}^{f,P}  \left(\bar\nu_\alpha \gamma_\mu P_L \nu_\beta\right) \left(\bar f \gamma^\mu P f\right) ~, 
\end{equation}
would lead to new NC interactions with the rest of the SM
fermions. Here, $\alpha,\beta\equiv e, \mu, \tau$ while $f$ refers to
SM fermions, and $P$ can be either a left-handed or a right-handed
projection operator ($P_L$ or $P_R$, respectively).  Such new
interactions may induce lepton flavour-changing processes (if $\alpha
\neq \beta$), or may lead to a modified interaction rate with respect
to the SM result (if $\alpha = \beta$).

As for their effect in CE$\nu$NS,
in presence of NC NSI, the CE$\nu$NS cross section can still be written as Eq.~\eqref{eq:xsec-SM}
with and  effective  charge of the nucleus  Eq.~\eqref{eq:QWSM}, ${\cal Q}_{W,\alpha}^2
\to {\cal Q}^2_\alpha(\Eps)$. For real NSI parameters, it can be written as ~\cite{Barranco:2005yy}
\begin{equation}
{\cal Q}^2_{\alpha}(\Eps) =  \left[ \mathcal{Q}_{W,\alpha} + 2\, \Eps_{\alpha\alpha}^\text{X} \right]^2 +4 \sum_{\beta\neq \alpha} \big( \Eps_{\alpha\beta}^\text{X} \big)^2 ~,
\label{eq:Qalpha-nsi-X}
\end{equation}
with
\begin{equation}
\Eps_{\alpha\beta}^\text{X} \equiv (N+2Z) \Eps^{u,V}_{\alpha\beta}
+ (2N+Z)\Eps^{d,V}_{\alpha\beta}
\label{eq:eps-nucleon}
\end{equation}
being the effective nuclear NSI vector couplings in terms of those of the
quarks, $\Eps^{q,V}_{\alpha\beta}\equiv \Eps^{q,L}_{\alpha\beta}+\Eps^{q,R}_{\alpha\beta}$.
Thus, the first consequence we observe of including NSI effects is
that the weak charge may now strongly depend on the incident neutrino flavour
$\alpha$.

At present, the best constraints available in the literature for these
operators come from global fits to oscillation data, which are very
sensitive to modifications in the effective matter potential felt by
neutrinos as they propagate in a medium
\cite{Coloma:2023ixt}. Consequently, they can strongly bound vector
NSI, and, since they are due to a totally coherent effect, these
bounds extend to NSI induced even by ultra light mediators ($M_{\rm
  med}\gtrsim 1/R_{\rm Earth}\sim {\cal O}(10^{-12})\,{\rm eV}$).
However, while oscillation experiments are sensitive to all
flavour-changing NSI, they are only sensitive to differences between
the diagonal NSI parameters in flavour space
\cite{Coloma:2023ixt,Coloma:2017egw}.  This leads to the appearance of
new degeneracies involving standard oscillation parameters and NSI
operators, such as the so-called LMA-DARK/generalized mass ordering
degeneracy~\cite{Miranda:2004nb,Gonzalez-Garcia:2013usa,Coloma:2016gei}.
Conversely, as seen above,  CE$\nu$NS experiments at spallation sources
allow to constrain two of the three flavour-diagonal coefficients,
lifting the degeneracy and allowing for the independent determination
of all the NSI with the combined analysis of oscillation and CE$\nu$NS
~\cite{Esteban:2018ppq,Coloma:2019mbs,Coloma:2023ixt}.  

With this in mind, in what follows we focus on the
determination of the flavour-diagonal NSI coefficients,
$\epsilon_{\alpha\alpha}^{f,V}$ ($f=u,d$), although it should be kept
in mind that coherent neutrino scattering is also sensitive to all the
off-diagonal NSI operators as well, and competitive sensitivity should also
be expected for those.

Our results on the expected sensitivity for the NSI coefficients are shown
Fig.~\ref{fig:NSI}. There we plot the 90\% C.L. allowed regions in the plane
$(\epsilon_{ee}^{u,V} ,\epsilon_{\mu\mu}^{u,V})$ for different detector 
materials (left panel). 
In this figure, for simplicity, we have assumed that the NSI take place
only with up-type quarks; however, similar results are obtained if the
NSI are assumed to take place with down-type quarks instead. 

Non-surprisingly, comparing Fig.~\ref{fig:NSI} and Fig.~\ref{fig:rsq}
we observe that the main qualitative features are very similar. In brief,
the allowed region consists of four distinct sections because ${\cal Q}_e^2$ 
and ${\cal Q}_{\mu}^2$ can be independently determined from timing information.
This results in two allowed ranges for each of the NSI coefficients considered. 
One of the four regions is centered at the SM prediction, 
($\epsilon_{ee}^{u,V} =\epsilon_{\mu\mu}^{u,V}=0$); the other three correspond to
the quasi degenerated solutions which in this case happen when 
$\Eps_{\alpha\alpha}^X=-{\cal Q}_{W,\alpha}$, i.e $\Eps_{\alpha\alpha}^{u,V}\simeq
\frac{N-Z(1-4\sin^2\theta_W)}{N+2Z}$. As in Fig.~\ref{fig:rsq}, for the Ar TPC, 
limited time resolution and large backgrounds hinder a separate measurement of ${\cal Q}_e^2$ 
and ${\cal Q}_{\mu}^2$. It is also clear from the figure 
that, because the impact of NSI on the weak charge depends on the values of $N$ and $Z$ in a
non-trivial manner, the combination of data obtained
for different nuclei offers an additional handle to reduce the size of
the allowed confidence regions of this scenario (for earlier
discussions see, e.g.,
Refs.~\cite{Scholberg:2005qs, Barranco:2005yy,Coloma:2017egw, Baxter:2019mcx}).

In the figure we also show for comparison the current constraints
  from COHERENT, as obtained in Ref.~\cite{DeRomeri:2022twg} by the
  combined analysis of results with CsI~\cite{COHERENT:2021xmm} and
  Ar~\cite{COHERENT:2020iec} targets. The figure clearly illustrates
  the improved sensitivity due to the flavour discrimination induced
  by the precise time-structure information of the neutrino signal
  stemming form the pulsed nature of the J-PARC proton beam, together
  with the significantly improved statistics.
  
We find that the effect of the different systematics and time information is also very similar
to that shown in the right panel of Fig.~\ref{fig:rsq}, i.e.,  the dominant source of systematic
uncertainty is the overall flux normalization. And most importantly,  as discussed above,
in absence of time information, the experiments are mostly sensitive to the combination 
${\cal Q}^2_e + 2 \, {\cal Q}^2_\mu$. This would result in a ring-like allowed
region around $(R + \epsilon_{ee}^{u,V} )^2 + 2 ( R + \epsilon_{\mu\mu}^{u,V})^2 = 3 R^2$
where now $R \equiv \frac{-N+Z (1-\sin^2\theta_W)}{2(2Z + N)}$. 
In this case, the only flavour information comes from the marginal variation of the
flavour composition with recoil energy.

As for the neutron radius measurement, the determination of NSI is not only statistics-limited.
As an illustration, if the CsI detector mass were increased up to 1 tonne (assuming that backgrounds
scale up proportionally), we find that the major axis of each allowed ellipse in Fig.~\ref{fig:NSI} would not change, 
as it is dominated by flux uncertainties (see Fig.~\ref{fig:rsq} right). The minor axis would, however,
shrink by a significant factor.

\subsection{Light vector mediators}
\label{sec:zp}
\begin{figure*}[ht!]
\includegraphics[width=0.45\textwidth]{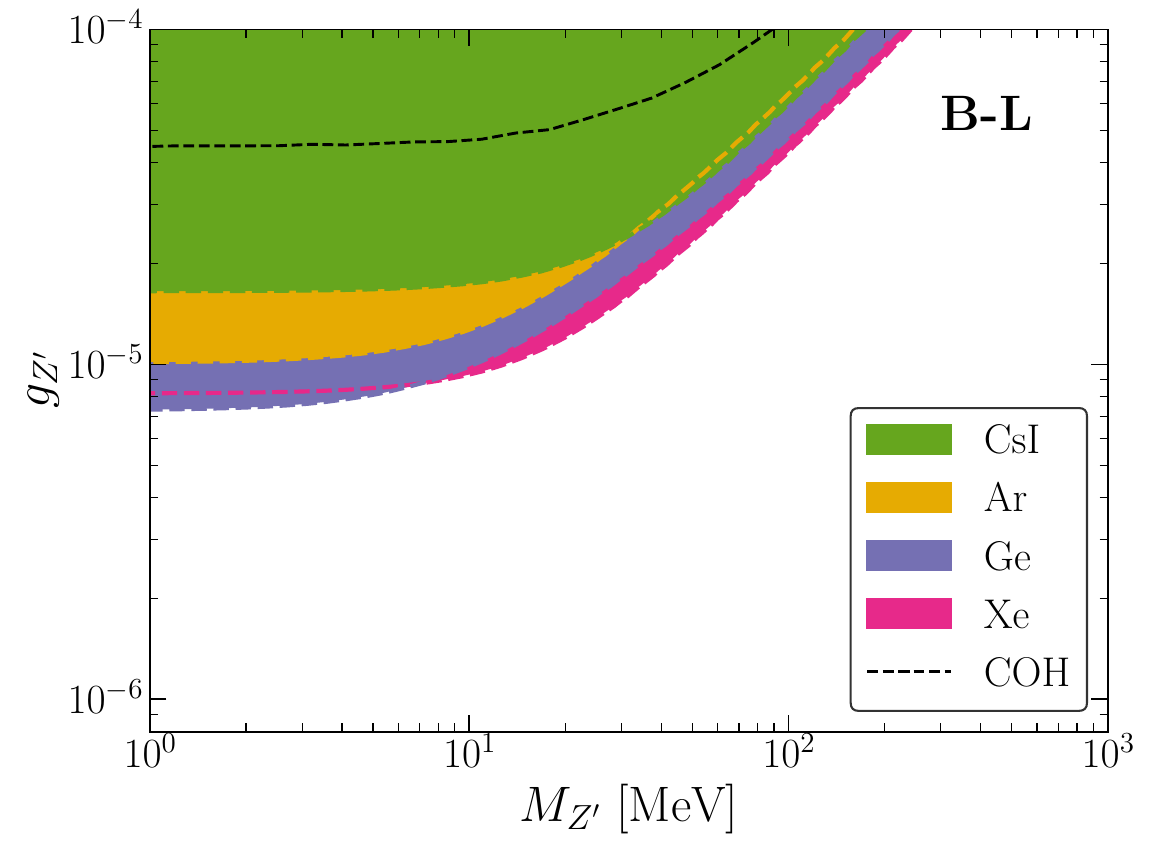}
\includegraphics[width=0.45\textwidth]{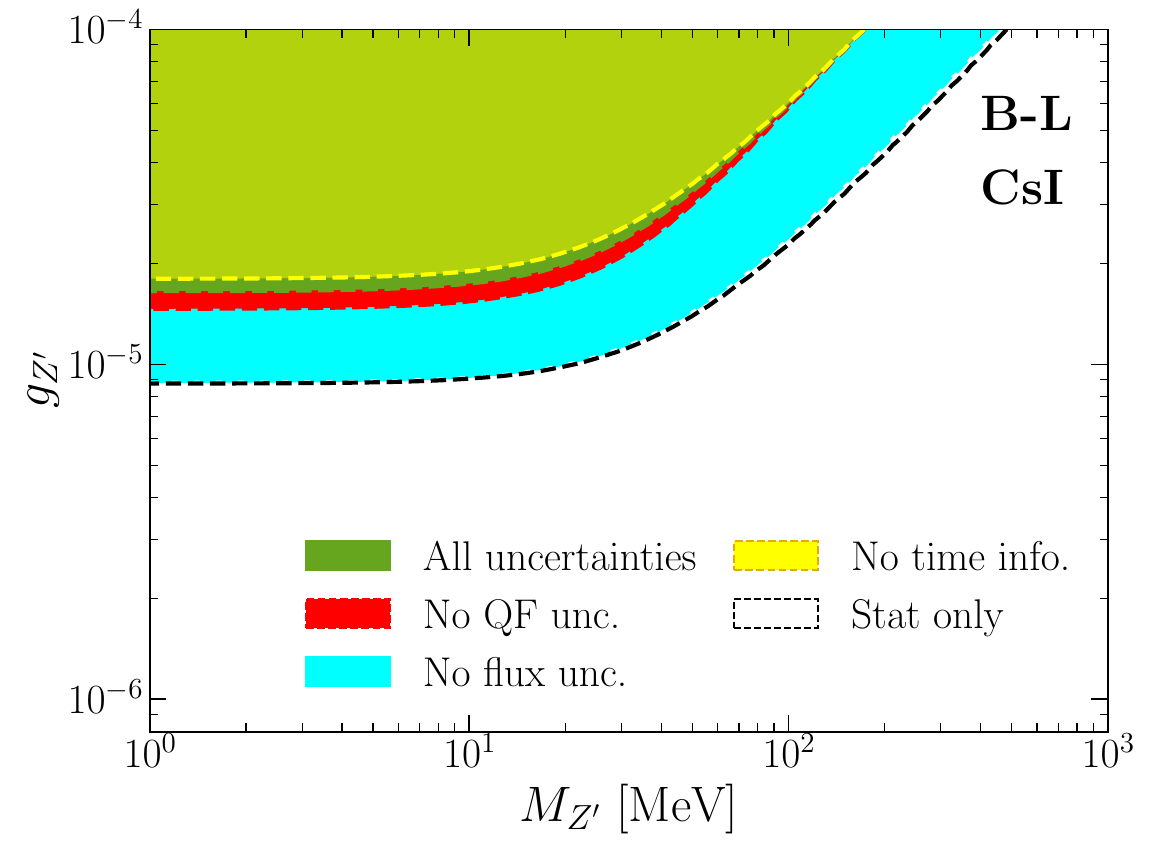}
\caption{Expected excluded regions in the
  $(g^{B-L}_{Z'} ,M_{Z'})$ plane at the 90\%
  confidence level (C.L.) for two d.o.f. ($\Delta\chi^2=4.61$).
 {\bf Left:} The  different regions correspond to the expected results for the
  different detectors listed in Table~\ref{tab:detectors}, as
  indicated by the legend.
  In all cases, the simulated data has been
  generated for the SM, and the results are then fitted assuming arbitrary
  values of the $Z'$ coupling and mass  and including the nominal
  uncertainties and time resolution. The dashed line shows the current 90\% CL bounds from the analysis of COHERENT CsI and Ar data from
  Ref.~\cite{DeRomeri:2022twg}.
{\bf  Right:} The excluded regions for the Cryogenic-CsI detector under different assumptions
  for the systematic uncertainties as labeled, see text for details. }
\label{fig:ZP}
\end{figure*}

$U(1)$ extensions of the SM gauge symmetry are one of the most minimal
forms of BSM physics. They appear in many top-down models which
contain a heavy ($\sim$ TeV scale) neutral gauge boson 
($Z'$)~\cite{Langacker:2008yv}. More recently, $U(1)$ extensions where new
gauge boson is light have been widely studied, in particular in
scenarios able to provide novel Dark Matter candidates~\cite{Fabbrichesi:2020wbt}.

From the point of view of their effect in CE$\nu$NS, these extensions
can be described in terms of the Lagrangian for the interaction of the
neutral vector $Z'$ with the fermions participating in the process:
\begin{equation}  
  {\cal L}_{Z'} = g_{Z'} Z'_\mu \!\!\!\sum_{\substack{f=u,d,e,\\\nu_{eL},\nu_{\mu L}}}
  \!\!\!\!\! q_{Z'}^f \left(\overline{f} \gamma^\mu f\right) +
  \frac{1}{2} M^2_{Z'} {Z'}^\mu Z'_\mu ~,
\end{equation}
where $q_{Z'}^f$ indicates the charge of each fermion $f$ under the
new $U(1)$ interaction.

The amplitude for neutrino scattering off nuclei mediated by the $Z'$
interferes with that of the SM, so the additional contribution to the
neutrino-nucleus scattering cross section reads~\cite{Cerdeno:2016sfi}
\begin{eqnarray}
  \Delta \frac{d \sigma_{Z',\alpha}}{d T} &=& \frac{g^2_{Z'} M}{2 \, \pi}
  \Big[\frac{g^2_{Z'} (q_{Z'}^\nu)^2 \mathcal{Q}^2_{Z'}}{\left(2 M T +
      M_{Z'}^2\right)^2} -\frac{\sqrt{2} G_F q_{Z'}^{\nu_\alpha}
      \mathcal{Q}_{Z'}\mathcal{Q}_{W}}{\left(2 M T + M_{Z'}^2\right)}
    \Big] \nonumber \\ &&\times \left(1-\frac{M T}{2 E^2_\nu}\right)
  |F(Q)|^2 ~,
\label{eq:csvecn}
\end{eqnarray}
where $\mathcal{Q}_{Z'}$ is the weak charge of the nucleus for the light 
vector interaction and $F(Q^2)$ is the form factor that, with enough precision at the
momentum transfers of interest, can be assumed to be the same as in the
SM~\cite{Hoferichter:2020osn}. Vector current
conservation implies that only valence quarks contribute by simply
summing up their charges, so for universal couplings ($q_{Z'}^u =
q_{Z'}^d\equiv q_{Z'}^q$), $\mathcal{Q}_{Z'} = 3 \, q_{Z'}^q \, (Z+N)$
(see, e.g., Ref.~\cite{DelNobile:2013sia}).

With existing CE$\nu$NS data, limits on a variety of vector boson
mediator models have been derived (see e.g. Refs.
~\cite{Liao:2017uzy,Papoulias:2017qdn,Coloma:2017ncl,Billard:2018jnl,Papoulias:2019txv,Khan:2019cvi,Giunti:2019xpr,Flores:2020lji,Cadeddu:2020nbr,AtzoriCorona:2022moj,AtzoriCorona:2022qrf,Coloma:2022avw,Liao:2022hno,DeRomeri:2022twg}
for an incomplete list).

For the sake of concreteness we show our results of the expected sensitivity
for a $Z'$ coupling to $B-L$ (for which $-1=q_{Z'}^{\nu_e=}=q_{Z'}^{\nu_\mu}=-3q_{Z'}^u=-3q_{Z'}^d$). This is an anomaly-free model which represents an example of a
fully consistent UV completion for the NSI phenomenological parametrization.
The results are shown in Fig.~\ref{fig:ZP}, where in the left panel we plot the
expected excluded regions in the $(g^{B-L}_{Z'} ,M_{Z'})$ plane
at the 90\% C.L. with the detectors under consideration.
For light mediator masses ($M_{Z'}\lesssim 5$--$10$ MeV depending on the detector
threshold), the experiment has no sensitivity to the mediator mass and the limit of
the regions approaches a horizontal line of constant coupling.
For these effectively massless scalar mediators,
it is possible to derive an upper bound on the coupling constant
independent of mediator mass, which we list in Table ~\ref{tab:macrotable}.
Conversely, for sufficiently large mediator masses, the boundary is a diagonal,
characteristic of the contact-interaction limit.
In that case, the event rates depend on $(g_{Z'}/M_{Z'})^2$. This is the limit
in which the effect can be described via an NSI parametrization
\begin{equation}
  \Eps^{u,B-L}_{\alpha\alpha}=\Eps^{d,B-L}_{\alpha\alpha}
  =\frac{1}{3\sqrt{2}G_F}\frac{g_{Z'}^2}{M_Z'^2}
\end{equation}  
We list in Table ~\ref{tab:macrotable} the bounds on $\frac{g_{Z'}^2}{M_Z'^2}$
in this regime.

For comparison, in the curve labeled COH, we show the current
  constraints from COHERENT, as obtained in
  Ref.~\cite{DeRomeri:2022twg} by the combined analysis of results
  with CsI~\cite{COHERENT:2021xmm} and Ar~\cite{COHERENT:2020iec}
  targets. (Consistent results but with a different CL were obtained
  in Ref.~\cite{AtzoriCorona:2022moj}.) As seen in the figure, for all
  detectors considered, one foresees an improvement with respect to
  current bounds from COHERENT. For example, the reach on $g_{Z'}$ can
  improve by a factor $\sim 3$--$5$ in the light-mediator mass regime.
  Light-mediator models are constrained by a plethora of laboratory
  experiments and well as cosmological observations. Comparing with
  the latest compilation in Fig.~5 in
  Ref.~\cite{AtzoriCorona:2022moj}, for the $B-L$ model CE$\nu$ES
  bounds provide the strongest constraints in the mass window
  $50\,{\rm MeV}\lesssim M_{Z'}\lesssim 500\,{\rm MeV}$.

The role of the different sources of uncertainty in this scenario
is shown in the right panel
of Fig.~\ref{fig:ZP}
(for the Cryogenic-CsI detector as illustration).
As seen in the figure, comparing the red contour and the
green region, removing the QF uncertainty has a marginal effect that is most relevant
at low mediator masses, where the experiment threshold sets the
sensitivity. We also see that, for this model, the timing information has very small
impact. This is expected, as it is a lepton-flavour universal model. 
In this case, the dominant uncertainty is the overall flux normalization. As seen
in the figure, reducing it can lead to an improvement of the sensitivity by a factor \mbox{$\sim$ 2}.

Finally, let us notice that in deriving the sensitivity to this scenario we have only
considered the effect of scattering off the nucleus, and we have not included
the possible effect of scattering off electrons in the target. Therefore, our sensitivity
can be considered to be conservative.

\subsection{Neutrino magnetic moment}
\label{sec:mu}
\begin{figure*}[ht!]
\includegraphics[width=0.45\textwidth]{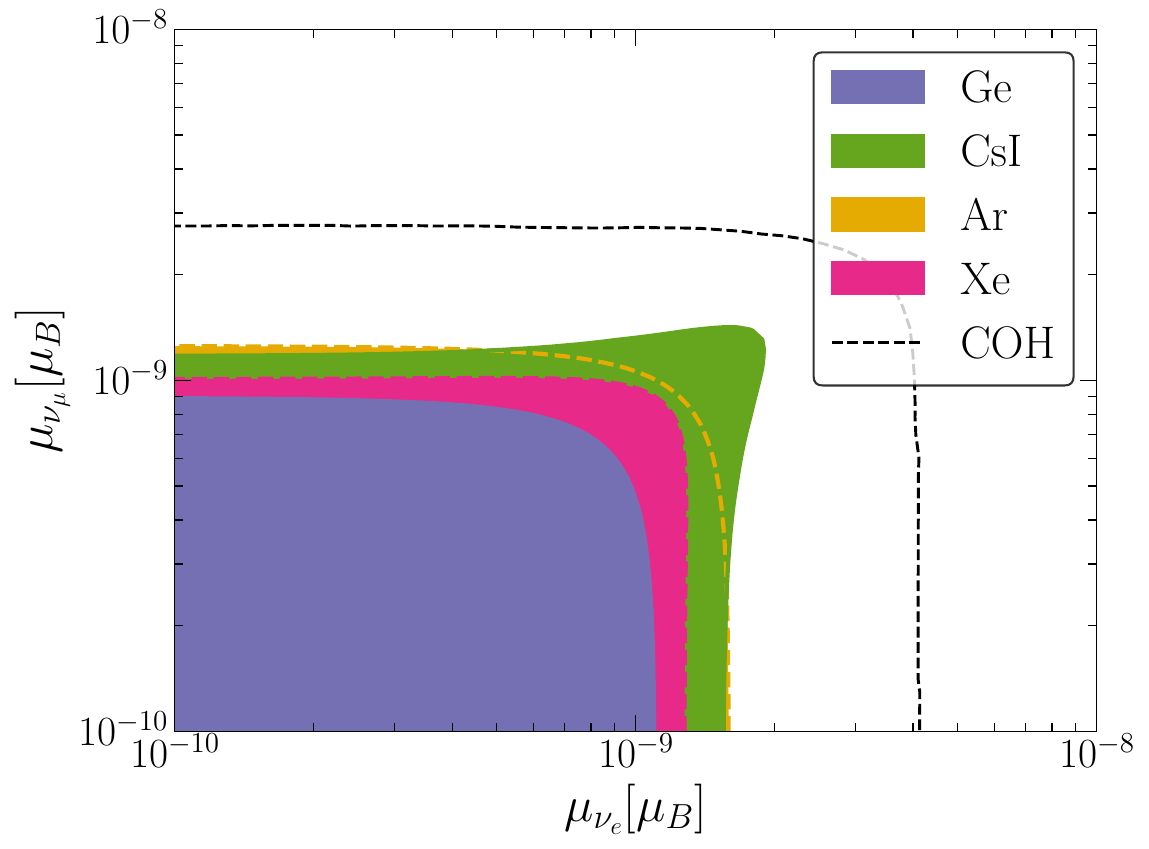}
\includegraphics[width=0.45\textwidth]{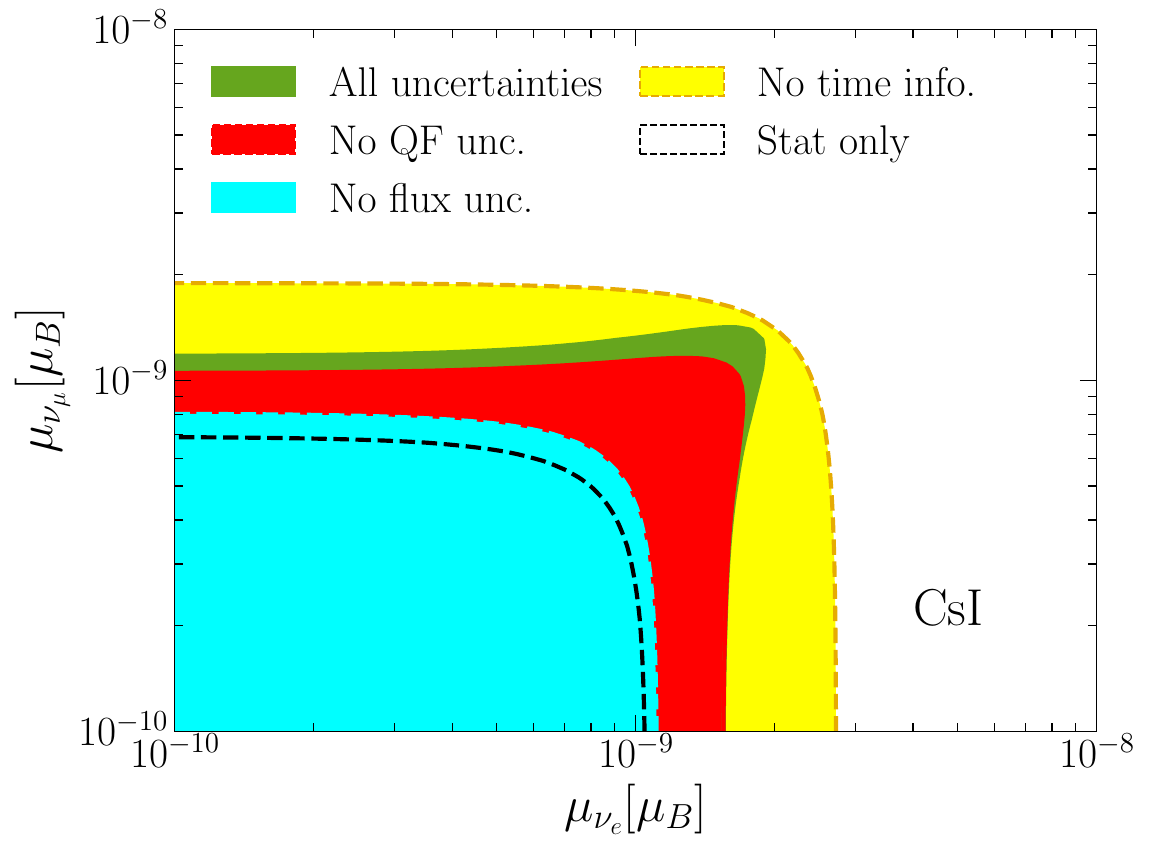}
\caption{Expected allowed regions in the
  $(\mu_{\nu_e} ,\mu_{\nu_\mu})$ plane at the 90\%
  confidence level (C.L.) for two d.o.f. ($\Delta\chi^2=4.61$).
  {\bf Left:} The  different regions correspond to the expected results for the
  different detectors listed in Table~\ref{tab:detectors}, as
  indicated by the legend.
  In all cases, the simulated data has been
  generated for the SM, and the results are then fitted assuming arbitrary values of the
  two magnetic moments coefficients and including the nominal uncertainties and time resolution.
  The dashed line shows the current 90\% CL bounds from the
    analysis of COHERENT CsI and Ar data from
    Ref.~\cite{DeRomeri:2022twg}.
  {\bf  Right:} The excluded regions for the Cryogenic-CsI detector under
  different assumptions
  for the systematic uncertainties as labeled, see text for details. }
\label{fig:MM}
\end{figure*}

The neutrino magnetic moment is the electromagnetic neutrino property
most experimentally searched for, as it is predicted to be non-zero
for massive neutrinos with non-negligible values in a wide spectrum of
BSM neutrino mass models. Laboratory and solar neutrino experiments
(among others) bound the $\nu_e$ ($\nu_\mu$) magnetic moments at the
level of $10^{-11}$ $\mu_{\rm B}$ ($10^{-9}$ $\mu_{\rm B}$)~\cite{ParticleDataGroup:2024cfk, Giunti:2024gec}.

In the presence of a neutrino magnetic moment, $\mu_\nu$, the cross
section for neutrino scattering off nuclei gets
additional contributions, which do not interfere with the SM ones. The
scattering off protons can be considered coherent and therefore its
cross section is given, up to order ${\cal O}(T/E_\nu)^2,
(T/M)$, by~\cite{Vogel:1989iv}
\begin{equation}
\label{eq:mag-nuc}	
\frac{d\sigma_{\rm \mu_\nu}}{d T} = Z^2 \,
\left(\frac{\mu_\nu}{\mu_B}\right)^2 \, \frac{\alpha^2\,\pi}{m_e^2}
\left[\frac{1}{T} - \frac{1}{E_\nu}\right]\; |F_{\rm em}(Q^2)|^2 \;.
\end{equation}
Since the effect due to a non-zero magnetic
moment is only relevant for very low recoils, the
electromagnetic form factor $F_{em}(Q^2)$ can be safely approximated
to one.
Notice also that, in writing Eq.~\eqref{eq:mag-nuc}, we have denoted the
neutrino magnetic moment as $\mu_\nu$, without specifying the neutrino
flavour. However, neutrino magnetic moments arise in a variety
of models of new physics and, in particular, they do not need to be
flavour-universal. Therefore, in what follows, we allow different
magnetic moments for the different neutrino flavours, reporting the
sensitivity separately.

Our results on the expected sensitivity to neutrino magnetic moments
with the detectors under consideration are shown in Fig.~\ref{fig:MM}
and summarized in Table~\ref{tab:macrotable}. From these expectations, we
see that the relative performance of the different detectors is
different than in the previous scenarios considered. In particular, 
we see that the Ar TPC sensitivity is comparable to that of the
CsI detector. This is so because the effect of a finite neutrino magnetic
moment on the
CE$\nu$NS event rates is most noticeable for small recoil
energies (below $0.5-1~{\rm keV }_{nr}$ for $\mu_{\nu_\mu} \sim
10^{-10}\, \mu_B$).  Therefore, it is expected that detectors with the
lowest possible recoil energy threshold will be more sensitive to this
neutrino property. Furthermore, the relative effect of the new
contribution compared to the SM grows with $\sim (Z/N)^2$ which is
$\sim$ 0.65 for Ar and Ge and $\sim$ 0.5 for Xe and CsI. 
Altogether, we find that the Xe TPC and HPGe PPC detectors  provide
the best sensitivity with a slightly better performance of the
HPGe PPC driven mostly by its lower threshold.

For comparison, we also show in the left panel the current
  constraints from COHERENT, as obtained in
  Ref.~\cite{DeRomeri:2022twg} by the combined analysis of results
  with CsI~\cite{COHERENT:2021xmm} and Ar~\cite{COHERENT:2020iec}
  targets. Equivalently, the quoted one-dimensional bounds read
  $\mu_{\nu_e(\nu_\mu)}<38 (26)\times 10^{-10}\, \mu_B$ at 90\% CL,
  which compared with the corresponding values quoted in
  Table~\ref{tab:macrotable} show an improvement of up to a factor
  ${\cal O} (3.5)$. For $\mu_{\nu_\mu}$ these bounds are competitive
  with constraints from other laboratory experiments, but they are
  always much weaker than those obtained from solar neutrinos in
  Borexino~\cite{Borexino:2017fbd,Coloma:2022umy} and in direct Dark
  Matter detection experiments (see the comprehensive compilation in
  Ref.~\cite{Giunti:2024gec}).

The role of the different sources of uncertainty and of the time information
in this scenario is further illustrated  in the right panel of Fig.~\ref{fig:MM}
(for the case of Cryogenic-CsI detector as an example). As we see, different uncertainties have
a comparable impact, with flux uncertainty being the most relevant. Time information is also 
relevant, particularly when the magnetic moment is flavor-dependent.

\subsection{Light sterile neutrinos}
\label{sec:ste}
\begin{figure*}[ht!]
\includegraphics[width=0.45\textwidth]{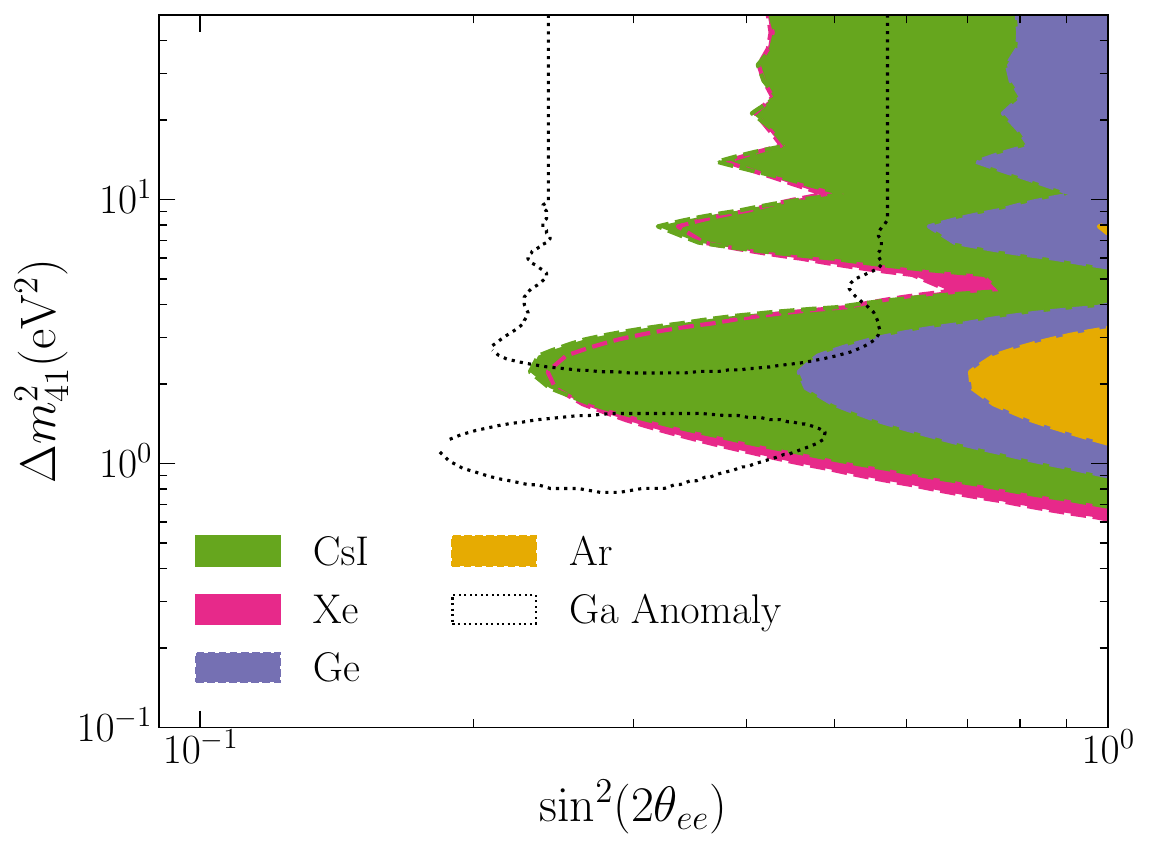}
\includegraphics[width=0.45\textwidth]{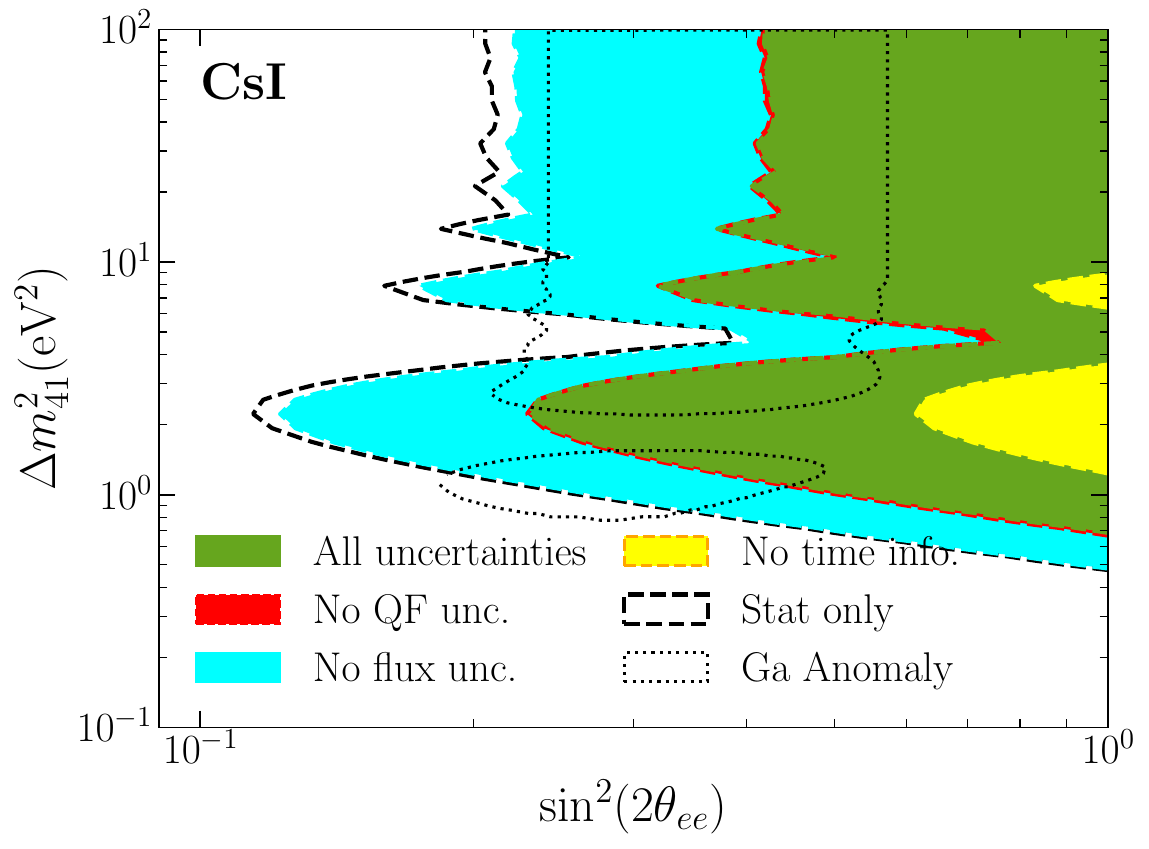}
\includegraphics[width=0.45\textwidth]{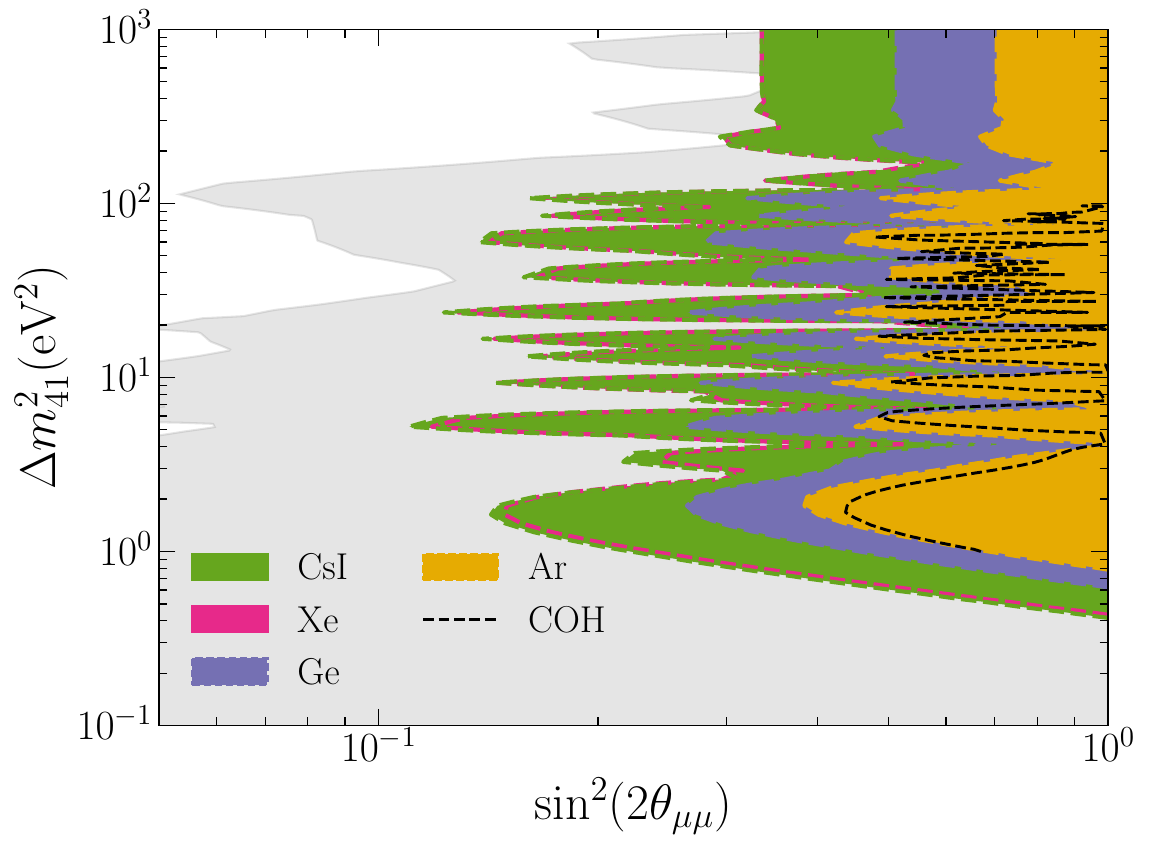}
\includegraphics[width=0.45\textwidth]{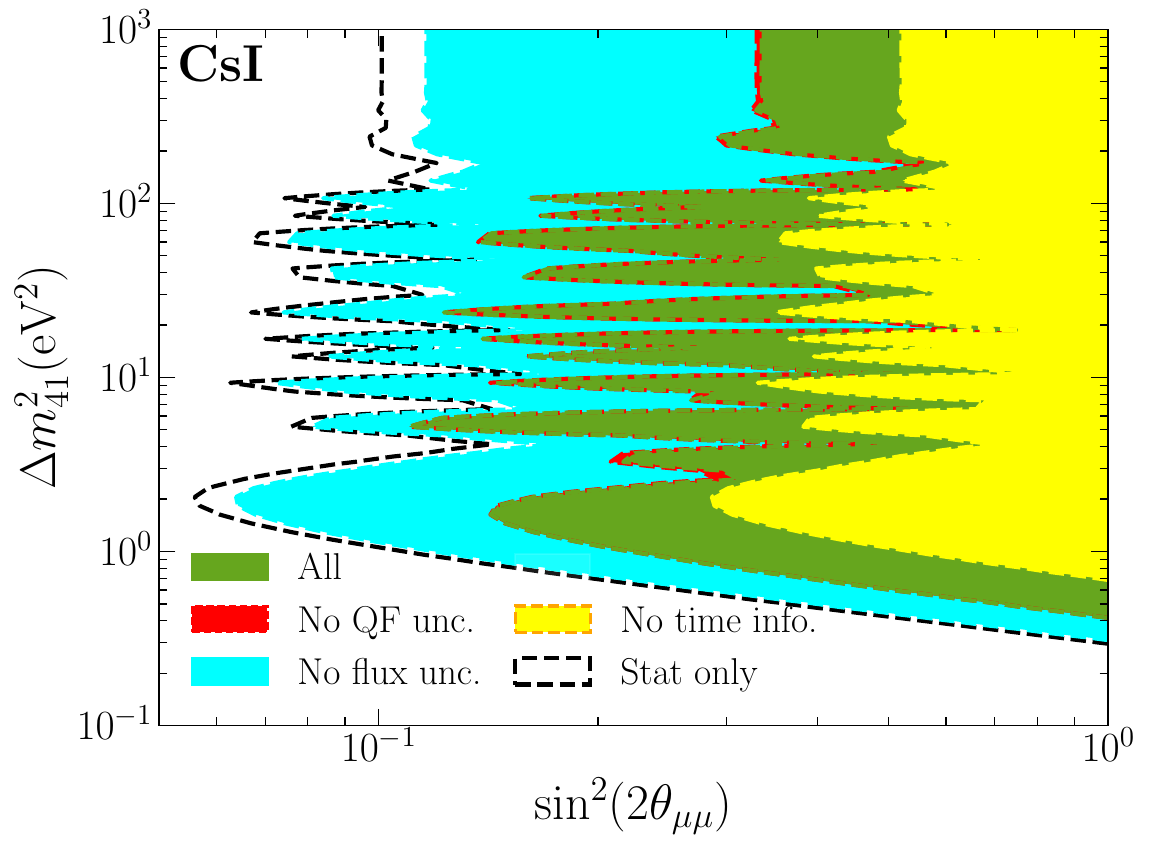}
\caption{Expected excluded regions in the
  $(\sin^2 2\theta_{ee(\mu\mu)}, \Delta m^2_{41})$ plane at the 90\%
  confidence level (C.L.) for two d.o.f. ($\Delta\chi^2=4.61$).
  {\bf Left:} The  different regions correspond to the expected results for the
  different detectors listed in Table~\ref{tab:detectors}, as
  indicated by the legend.
  In all cases, the simulated data has been
  generated for the SM, and the results are then fitted assuming arbitrary values of the
  two oscillation parameters and including the nominal uncertainties and time resolution. 
{\bf  Right:} The excluded regions for the Cryogenic-CsI detector under different assumptions
  for the systematic uncertainties as labeled, see text for details. 
  In the upper plots for comparison we also show the
  2$\sigma$ region of the parameters which could explain the Gallium anomaly (from Ref.~\cite{Barinov:2021asz}). In the lower left plot
  we show the bounds from the analysis
  of COHERENT CsI and Ar data from Ref.~\cite{DeRomeri:2022twg} (dashed lines),
  and the envelope of current limits from the most constraining experiments in this
  mass range~\cite{NOvA:2024imi, MINOS:2020iqj, SciBooNE:2011qyf, Stockdale:1984cg}  (shaded grey region).}
\label{fig:sterile}
\end{figure*}

Historically, models with extended light neutrino sectors were invoked
to explain a set of anomalies observed at short baselines that could 
not be explained within the standard 3$\nu$ framework. These anomalies
could be interpreted as hints for the existence of additional
neutrino states with masses at the eV scale. This requires the introduction
of sterile neutrinos, i.e., SM singlets that mix with the three standard neutrinos,
generating flavour oscillations at shorter distances.

At the baselines $\mathcal{O}(m)$ considered for CE$\nu$NS, the dominant effect of  
such oscillations is the depletion of active neutrino fluxes arriving
to the detector. As CE$\nu$NS is sensitive to all neutrino flavors with the same
SM interactions, it can directly test this depletion.
Thus, it has been studied to provide constraints on this scenario,
which can be relevant for the interpretation of the SBL
anomalies~\cite{Blanco:2019vyp,Miranda:2020syh,Bisset:2023oxt}.

Focusing on the 3+1 scenario, the expected number of events can be obtained 
as described in Sec.~\ref{sec:param} (neglecting the effect of the neutrino charge
radius), where now the energy dependent function is 
\begin{equation}
  \!\!\!\!  g_{X}(T) \propto \int_{E_\nu^\mathrm{min}}^\infty 
  \!\!\!\!\!\!\!\!dE_\nu \, P_{Xa}(E_\nu)\,\phi_X(E_\nu) \frac{d\sigma_X}{dT} (E_\nu, T) \, ,
\end{equation}
with the probabilities of neutrinos of flavour $X = \{e, \mu\}$ converting into \emph{any} active neutrino given by
\begin{equation}
  P_{Xa}(E_\nu)=1-4|U_{X4}|^2(1-\!\!\!\!\sum_{\alpha=e,\mu,\tau}\!\!\!\!|U_{\alpha 4}|^2)
  \sin^2\left(\frac{\Delta m^2_{41}}{2 E_\nu}\right) \;.
\end{equation}

For concreteness, we consider two cases.  In the first case we
  will assume $U_{\mu 4}=U_{\tau 4}=0$, for which $P_{\mu a}$=1 and
\begin{equation}
P_{ea}=1-\sin^22\theta_{ee} \sin^2\left(\frac{\Delta m^2_{41}}{2
  E_\nu}\right)
\end{equation}
with $\sin^2\theta_{ee}\equiv|U_{e4}|^2$. This is the scenario invoked as
possible explanation for the so-called Gallium
anomaly~\cite{Acero:2007su,Barinov:2021asz}. In brief, the radioactive 
source experiments at the Gallium
solar neutrino experiments both in SAGE and GALLEX/GNO obtained an
event rate lower than expected. This result was confirmed by the BEST
experiment~\cite{Barinov:2021asz}.  The effect can be explained by the
hypothesis of $\nu_e$ disappearance due to oscillations with $\Delta
m^2 \gtrsim 1~$eV$^2$ ~\cite{Acero:2007su}. This interpretation,
however is in tension with the analysis of solar neutrino
experiments~\cite{Gonzalez-Garcia:2024hmf}, and with latest KATRIN
results~\cite{KATRIN:2025lph}. Interestingly, CE$\nu$NS can directly 
test the hypothesis of neutrinos converting into a flavour without SM 
neutral current interactions, with less dependence on the underlying mechanism.

In the second  case we assume  $U_{e 4}=U_{\tau 4}=0$, for
which $P_{e a}$=1 and
\begin{equation}
P_{\mu a}=1-\sin^22\theta_{\mu\mu} \sin^2\left(\frac{\Delta m^2_{41}}{2
  E_\nu}\right)
\end{equation}
with $\sin^2\theta_{\mu\mu}\equiv|U_{\mu4}|^2$.

We plot in Fig.~\ref{fig:sterile} the regions of oscillation parameters
which can be excluded at the 90\% C.L. with the detectors under consideration.
In the upper panels we show the results for the $U_{\mu 4}=U_{\tau 4}=0$  case. 
For sake of comparison, we also show the
2$\sigma$ parameter region which could explain the anomaly~\cite{Barinov:2021asz}.
As seen in the figure, the best sensitivity for this scenario is
attainable with the Cryogenic-CsI and the Xe TPC detectors
 which can probe part of the parameter space
  required to explain the anomaly, though they are not competitive with
  the bounds from solar experiments and KATRIN..
In the right panel, we show the impact of the different   uncertainties and
of the time information in this scenario on this conclusion. As seen in the figure, the dominant
uncertainty is that of the overall flux normalization, which is expected since
the effect of the oscillation into sterile neutrinos is in this case the
depletion of the $\nu_e$ flux. Since $\nu_\mu$s are not affected, the sensitivity
is strongly dependent on flavour discrimination, hence the relevance of
the timing information observed in the right panel of the figure. 

Conversely, the lower panels show the results for the $U_{e4}=U_{\tau 4}=0$ case. Because of the monochromatic nature of prompt $\nu_\mu$,
  the oscillatory behaviour of the survival probability does not get averaged
  by the experimental energy resolution. In this case, we consider the
  effect of the smearing over the uncertainty in the baseline, which for
  an ${\cal O}( \%)$ uncertainty results into averaged oscillations for
  $\Delta m^2_{41}\sim {\cal O}({\rm few}\, 10^{2})$ eV$^{2}$. 
  In our analysis, we adopt a $0.5\%$ baseline
	uncertainty for all targets.
	We have verified that varying this uncertainty from 1~cm to 30~cm
	does not affect the exclusion contour at low masses
	($\Delta m^2_{41} \lesssim 100$~eV$^2$), and it only slightly modifies
	the amplitude of the last two oscillation peaks before averaging it out.

For this channel, we can
  also compare with the sensitivity obtained with the analysis of
  COHERENT CsI and Ar data in Ref.~\cite{DeRomeri:2022twg}, over which
  we find that all detectors can provide an improvement.
  We also compare
  with the present exclusion from the most constraining experiments in this
  mass range ~\cite{NOvA:2024imi,MINOS:2020iqj,SciBooNE:2011qyf,Stockdale:1984cg}  shown as the shaded grey region in the figure.
  As seen in the figure, some of the detectors in consideration hold the
  potential to yield dominant sensitivity for
  $\Delta m^2_{41}\sim (2$--$4)\times 10^{2}$ eV$^2$ provided the aforementioned uncertainty
  on the baseline can be guaranteed. Finally, from the right panels we see
  that the relative behaviour of the different
  detectors and the relevance of the impact of the different
  uncertainties is similar for both oscillation channels.

\section{Conclusions}
\label{sec:conclusions}

In this work we have explored the prospects of J-PARC MLF for CE$\nu$NS
studies.  In order to quantitatively assess the potential of our proposal, we have explored the sensitivity
to a broad spectrum of SM effects ---including the weak mixing angle, the neutrino charge
radii and the neutron nuclear radius --- and BSM scenarios, such as NSI, new light mediators,
neutrino magnetic moments, and $\mathcal{O}$(eV) sterile neutrinos. We have considered as examples a
variety of detector technologies presently funded for construction and under development, with
different nuclear targets (listed in Table~\ref{tab:detectors}). In our studies, we have taken into account the careful
characterization and evaluation of the expected background sources and their energy and time
dependence, as described in Sec.~\ref{sec:bckg_sources}.

We have also analyzed the relative effect of the main sources of systematic uncertainties: the
overall flux normalization, the QF uncertainty, and the uncertainty on the normalization of the
irreducible backgrounds. Our results show that for all scenarios and detector
technologies considered, the overall flux normalization is the dominant systematic uncertainty.
Any improvement on this uncertainty would largely enhance the sensitivity of the expected vast
statistics (see, e.g., Fig.~\ref{fig:speccsi}).

We also find that for flavour-independent effects, such as variations of the weak mixing angle,
or scenarios with new universal interactions, all technologies deliver comparable sensitivity.
Differences are only determined by the statistics attainable in the different detectors, and/or the
nuclear charge dependence of the scenario (see Table~\ref{tab:macrotable} and Fig.~\ref{fig:ZP}). Conversely, larger differences in the
  performance are found for flavour-dependent effects (in particular
  for charge radii, NSI, and sterile neutrinos). For them,
  including the precise time-structure information of the neutrino 
  signal stemming form the pulsed nature of the J-PARC proton beam
  substantially boosts the sensitivity for detectors with good time 
  resolution (see Figs.~\ref{fig:rsq} and
  ~\ref{fig:sterile}). A notable exception to this general behavior is
  the neutrino magnetic moment, for which the reach of a given
  detector also depends strongly on its recoil energy threshold.  All
  in all, our results illustrate the well--known boost in
  sensitivity attainable by employing a variety of detectors with different
  nuclear targets. For all scenarios studied we have
  showed the improvement in sensitivity compared with current
  constraints from the from the observation of CE$\nu$NS at the
  spallation source experiment COHERENT.  In particular, we find that
  the expected sensitivity can result in leading bounds on neutrino
  charge radius, neutron nuclear distribution radius, laboratory
  constraints on NC-NSI, interactions with
  light mediators in the mass window ${\cal O}$(100 MeV), and potentially
  $\nu_\mu$-sterile oscillations with
  $\Delta m^2 \sim {\cal O}({\rm few}\,\times 10^{-2}\,{\rm eV}^2)$.

The J-PARC proton beam already serves important neutrino experiments including
Super-Kamiokande and JSNS$^{2}$. As we have shown, CE$\nu$NS measurements are not only
feasible at this facility, but they would provide leading statistics and excellent physics sensitivity. As such, 
they would complement the physics program of Hyper-Kamiokande, for example,
by breaking well-known degeneracies between standard oscillation and
some New Physics effects in long baseline neutrino oscillation experiments. Altogether, CE$\nu$NS activities would significantly
boost the neutrino-physics output of J-PARC MLF, with relatively low effort.
  

\section*{Acknowledgments}
This work is funded 
 by a Basque Government grant IT1628-22, MICIU/AEI/10.13039/501100011033 grants PID2022-136224NB-C21, PID2021-123703NB-C21, PID2022-136510NBC33, PID2024-156016NB-I00, 
and CEX2024-001451-M, and by USA-NSF grants PHY-2210533 and PHY-2209579.
It has also received support from the
Horizon Europe research and innovation programme under the Marie Sklodowska-Curie
Staff Exchange grant agreement No 101086085 – ASYMMETRY.
This project has also been supported by the European Research Council (ERC) under Grant Agreements No. 101039048-GanESS and ERC Advanced Grant 101055120 (ESSCEvNS).
This project is also supported by the National Natural Science Foundation of China (12425506, 12375101, 12090060, and 12090064) and the SJTU Double First Class start-up fund (WF220442604). Part of this work used the Solaris cluster, acquired through the Basque Government IT1628-22 grant.

\bibliographystyle{JHEPmod}
\bibliography{bibliografia.bib}
\end{document}